\documentclass[12pt,reqno]{amsart}

\usepackage{amsmath,
amsthm,
cite}

\usepackage{fullpage}

\usepackage{times}

\usepackage{psfrag}
\usepackage[dvips]{graphicx}
\numberwithin{equation}{section}
\newtheorem{theorem}{Theorem}
\newtheorem{prop}[theorem]{Proposition}
\newtheorem{coro}[theorem]{Corollary}

\newtheorem{conj}{Conjecture}

\DeclareMathOperator{\sh}{sh}
\DeclareMathOperator{\ch}{ch}
\DeclareMathOperator{\erfc}{erfc}

\usepackage{titlesec}
\titleformat{\section}
{\bfseries\scshape\centering}
{\thesection.}{.5em}{}
\titleformat{\subsection}
{\rmfamily\bfseries}
{\thesubsection}{.5em}{}

\begin{document}
\baselineskip 19pt
\parskip 7pt

\renewcommand{\thefootnote}{\fnsymbol{footnote}}

\sloppy




\title{
  Quantum  Invariant for Torus Link and
  Modular Forms
}


    \author{Kazuhiro \textsc{Hikami}}


  \address{Department of Physics, Graduate School of Science,
    University of Tokyo,
    Hongo 7--3--1, Bunkyo, Tokyo 113--0033, Japan.
    }

    \email{\texttt{hikami@phys.s.u-tokyo.ac.jp}}


\date{May 15, 2003. Revised on October 17, 2003.
  To appear in
  Commun. Math. Phys.}

\begin{abstract}
 We consider an asymptotic expansion of Kashaev's invariant
 or of the colored Jones function for the torus link $T(2,2 \, m)$.
 We shall give  $q$-series identity related to these invariants,
 and
 show that the invariant is regarded as
 a limit of $q$ being $N$-th root of unity of the Eichler integral of
 a modular form
 of weight
 $3/2$ which is related to the  $\widehat{su}(2)_{m-2}$ character.

\end{abstract}





\maketitle
\section{Introduction}

Recent studies reveal an intimate connection between
the quantum knot invariant
and  ``nearly modular forms'' especially with  half integral
weight.
In Ref.~\citen{LawrZagi99a} Lawrence and Zagier studied an asymptotic
expansion of the Witten--Reshetikhin--Turaev invariant of the
Poincar{\'e} homology sphere, and they showed that the invariant
can be regarded as the Eichler integral of a  modular form of weight
$3/2$.
In Ref.~\citen{DZagie01a},
Zagier further  studied a  ``strange identity''
related to the half-derivatives of the Dedekind
$\eta$-function, and
clarified a role of the Eichler integral
with half-integral weight.
{}From the viewpoint of the quantum invariant, Zagier's $q$-series
was originally connected with a generating function of
an upper bound of the number of linearly
independent Vassiliev invariants~\cite{AStoime98a},
and later it was
found that Zagier's $q$-series with $q$ being the $N$-th root of unity
coincides  with Kashaev's invariant~\cite{Kasha95,Kasha96b},
which was
shown~\cite{MuraMura99a} to coincide with a specific value of the colored
Jones function, for
the trefoil knot.
This correspondence was further investigated for the torus knot, and
it was  shown~\cite{KHikami02c}  that
Kashaev's invariant for the torus knot $T(2,2 \, m+1)$
also has a nearly modular property;
it can be regarded as a limit $q$ being the root of unity of
the Eichler integral of the Andrews--Gordon
$q$-series, which is theta series with weight $1/2$ spanning
$m$-dimensional space.
As the torus knot is not hyperbolic,
studies of the torus knot may not be attractive for the
``Volume  Conjecture''~\cite{Kasha95,MuraMura99a}
which states that an asymptotic limit of Kashaev's invariant coincides with the
hyperbolic volume of the knot complement,
but they are rather  absorbing from
the point of view of the number theory, $q$-series  and
modular forms.

\begin{figure}[htbp]
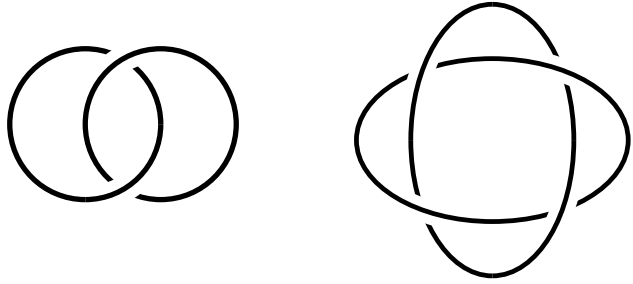

  \centering
  \begin{tabular}{cp{4cm}c}
    \includegraphics[scale=1, bb=10 -60 20 40]{hopf.ps}
    &
    &
    \includegraphics[scale=0.9, bb=0 -60 10 60]{knot24.ps}
  \end{tabular}
  \caption{Hopf link $T(2,2)$ and torus link $T(2,4)$}
  \label{fig:Hopf}
\end{figure}

Motivated by our previous result on the torus knot $T(2,2\,m+1)$,
we study Kashaev's invariant for the
torus link  $T(2,2 \,m)$
(see Fig.~\ref{fig:Hopf}) in this article.
We shall  show that the invariant is now regarded as the
\emph{half-integration} or the Eichler integral of a  modular form of
weight $3/2$.
Remarkable is that this modular form is related to the 
$\widehat{su}(2)_{m-2}$ character.
It is noted that recent studies~\cite{Zwege01a,Zweg02Thesis} reveal a
relation with Ramanujan's mock theta functions.
We also propose
a  $q$-series identity, which is new as far as we know,
and study an asymptotic expansion thereof.

This paper is organized as follows.
In section~\ref{sec:Jones} we construct the colored Jones polynomial
for the torus link $T(2,2 \,m)$.
Using the Jones--Wenzl idempotent, we give an explicit formula of the
invariant.
It is known~\cite{MuraMura99a} that
Kashaev's invariant coincides with a specific value of the colored
Jones polynomial.
This  correspondence enables us
to  give an integral form of Kashaev's invariant for the torus
link $T(2, 2\,m)$ in section~\ref{sec:Kashaev}.
We further give an asymptotic expansion of the invariant, and see that
the invariant for $T(2, 2 \, m)$ also has a  nearly modular property.
We  give an explicit form of Kashaev's
invariant for this torus link using the enhanced Yang--Baxter
operator.
Combining these results we obtain an asymptotic expansion of a
certain $\omega$-series.
In section~\ref{sec:q_series} we introduce the $q$-series related to
Kashaev's invariant for torus link,
and prove a
new $q$-series identity.
We study the modular property of these $q$-series, and discuss
how
Kashaev's invariant for $T(2,2\,m)$  may be  regarded as the Eichler
integral of a  modular form with weight $3/2$ which is the affine
$\widehat{su}(2)_{m-2}$ character,
in  section~\ref{sec:modular}.
In the last section, we  collect some examples.

\section{\mathversion{bold}
  Colored Jones Polynomial for Torus Link $(2, 2 \, m)$}
\label{sec:Jones}

The $N$-colored Jones polynomial for torus knot $T(m,p)$ was studied in
Refs.~\citen{Mort95a,RossJone93a}.
Following these methods,
we   compute the colored Jones polynomial for torus
link $T(2,2 \, m)$  in this section.
We use the Jones--Wenzl idempotent, and use following formulae
(see, \emph{e.g.}, Ref.~\citen{Licko97Book});
\begin{gather}
  \label{JW_1}
  \begin{psfrags}
    \psfrag{a}{$a$}
    \psfrag{b}{$b$}
    \psfrag{c}{$c$}
    \includegraphics[scale=0.5, bb=0 -40 60 80]{twist.ps}
  \end{psfrags}
  =
  (-1)^{(a+b-c)/2} \,
  A^{a+b-c+\frac{a^2 + b^2 - c^2}{2}} \,
  \begin{psfrags}
    \psfrag{a}{$a$}
    \psfrag{b}{$b$}
    \psfrag{c}{$c$}
    \includegraphics[scale=0.5, bb=-80 -40 80 80]{normal.ps}
  \end{psfrags}
  \\[2mm]
  \label{JW_2}
  \begin{psfrags}
    \psfrag{a}{$a$}
    \psfrag{b}{$b$}
    \includegraphics[scale=0.7, bb=-50 0 60 100]{fad.ps}
  \end{psfrags}
  =
  \sum_{c:
    \text{
      $(a,b,c)$ is admissible
    }}
  \frac{
    \Delta_c
  }{
    \theta(a,b,c)
  } 
  \begin{psfrags}
    \psfrag{a}{$a$}
    \psfrag{b}{$b$}
    \psfrag{c}{$c$}
    \includegraphics[scale=0.6, bb=-50 0 100 100]{boot.ps}
  \end{psfrags}
  \\[2mm]
  \nonumber
\end{gather}
where each label denotes a color, and we mean that
\begin{gather*}
  \Delta_n
  = (-1)^n \,
  \frac{A^{2 (n+1)} - A^{- 2 (n+1)}}{A^2 - A^{-2}} ,
  \\[2mm]
  \text{$(a,b,c)$ is admissible}
  \Leftrightarrow
  \text{$a+b+c$ is even, and}
  \begin{cases}
    a \leq b+c, \\
    b \leq c+a, \\
    c \leq a+b .
  \end{cases}
\end{gather*}
We   have a $\theta$-net
\begin{gather}
  \label{JW_theta}
  \theta(a,b,c)
  =
  \begin{psfrags}
    \psfrag{a}{$a$}
    \psfrag{b}{$b$}
    \psfrag{c}{$c$}
    \includegraphics[scale=0.6, bb=-60 -3 60 60]{theta.ps}
  \end{psfrags}
  \\[4mm]
  \nonumber
\end{gather}
which is given as
\begin{gather*}
  \theta(a,b,c)
  =
  \frac{
    \Delta_{x+y+z}! \, \Delta_{x-1}! \, \Delta_{y-1}! \, \Delta_{z-1}!
  }{
    \Delta_{y+z-1}! \, \Delta_{z+x-1}! \, \Delta_{x+y-1}!
  } ,
\end{gather*}
with
\begin{align*}
  a & = y + z,
  &
  b & = z + x,
  &
  c & = x+ y .
\end{align*}

\begin{prop}
  \label{prop:Jones_link}
  The $N$-colored Jones polynomial
  $J_N(h; \mathcal{K})$ for the torus link
  $\mathcal{K}=T(2,2 \, m)$ is given by
  \begin{equation}
    \label{Jones_link}
    2 \sh \left(\frac{N \, h}{2}\right) \,
    \frac{
      J_N(h;\mathcal{K})
    }{
      J_N(h;\mathcal{O})
    }
    =
    \mathrm{e}^{
      -\frac{1}{2} m (N^2 -1) h
    }
    \sum_{\varepsilon = \pm 1}    \sum_{j=0}^{N-1}
    \varepsilon \,
    \mathrm{e}^{
      m h j^2 + (m+ \varepsilon) \, h \, j
      + \frac{1}{2} h \varepsilon
    } ,
  \end{equation}
  where a parameter $q$ is set to be
  \begin{equation*}
    q= A^4 = \mathrm{e}^{h} ,
  \end{equation*}
  and $\mathcal{O}$ denotes unknot whose invariant is given by
  \begin{equation*}
    J_N(h;\mathcal{O})
    =
    \frac{
      \sh(N \, h/2)
    }{
      \sh (h/2)
    } .
  \end{equation*}
\end{prop}

\begin{figure}[htbp]
  \centering
  \begin{psfrags}
    \psfrag{N}{$n$}
    \psfrag{c}{$c$}
    \psfrag{K}{$\large{\Longrightarrow}$}
    \includegraphics{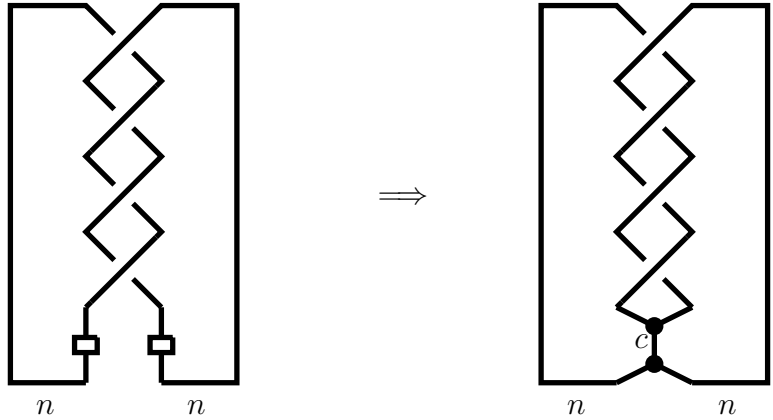}
  \end{psfrags}
  \caption{We apply eq.~\eqref{JW_2} to the torus link $T(2,4)$.
    We set  $n=N-1$.}
  \label{fig:jw}
\end{figure}

\begin{proof}
  We first  apply eq.~\eqref{JW_2} in the torus link $T(2,2 \, m)$
  (see Fig.~\ref{fig:jw}),
  and untangle crossings
  recursively using eq.~\eqref{JW_1}.
  We see that $\theta(a,b,c)$ vanishes at the end.
  We have
  \begin{align*}
    J_N(h; \mathcal{K})
    & =
    \sum_{c:
      \text{$(N-1,N-1,c)$ is admissible}
    }
    \Delta_c \,
    \left(
      (-1)^{N-1- \frac{c}{2}} \,
      A^{-2 (N-1) + c - (N-1)^2 + \frac{1}{2} c^2}
    \right)^{2 m}
    \\
    & =
    \frac{A^{-2 m (N^2 -1)}}{A^2 - A^{-2}}
    \sum_{j=0}^{N-1}
    A^{4 m j(j+1)} \,
    \left(
      A^{2 (2j+1)} - A^{-2 (2j+1)}
    \right) .
  \end{align*}
  This proves eq.~\eqref{Jones_link}.
\end{proof}

\section{Kashaev Invariant and Asymptotic Expansion}
\label{sec:Kashaev}

It is known~\cite{MuraMura99a}  that
Kashaev's  invariant is given from  the colored Jones
polynomial  at a specific value
$h \to 2 \, \pi \, \mathrm{i}/N$.
By use of a result of
Prop.~\ref{prop:Jones_link} we obtain an integral form of the
invariant as follows.

\begin{prop}
  The Kashaev invariant for torus link $T(2,2 \, m)$ is given by an
  integral form as
  \begin{multline}
    \label{link_integral}
    \langle T(2, 2 \, m)\rangle_N
    =
    \mathrm{e}^{
      \frac{\pi \mathrm{i}}{2 N} (m - \frac{1}{m})
    } \, 
    8 \sqrt{2} \, (m \, N)^{\frac{3}{2}} \, 
    \mathrm{e}^{-\pi \mathrm{i}/4}
    \\
    \times
    \int\limits_\mathcal{C} \mathrm{d} w \,
    w^2 \, 
    \mathrm{e}^{8 m \pi \mathrm{i} N w^2 + 4 m \pi N w}
    \,
    \frac{
      2 \, \sh(4 \, m \, \pi \, N \, w) \,
      \sh(4 \, \pi \, w)
    }{
      \sh(4 \, m \, \pi \, w)
    } ,
  \end{multline}
  where $\mathcal{C}$ denotes a path passing  through the origin in
  the steepest descent direction.
\end{prop}

\begin{proof}
  A proof is essentially same with those given in
  Refs.~\citen{KashaTirkk99a,KHikami02b,Rozan96a,LawreRozan99a} for studies in
  asymptotic behavior of the quantum invariants.

  To rewrite eq.~\eqref{Jones_link} into an integral form, 
  we use an integral formula
  \begin{equation*}
    \mathrm{e}^{h \, w^2}
    =
    \frac{1}{\sqrt{\pi \, h}}
    \int\limits_\mathcal{C} \mathrm{d} z \,
    \mathrm{e}^{- \frac{z^2}{h} + 2 w z} ,
  \end{equation*}
  where an integration path $\mathcal{C}$ is an infinite line passing
  through the   origin in the steepest descent direction.
  Then we get
  \begin{align*}
    2 \sh\left(\frac{N \, h}{2} \right) \,
    \frac{J_N(h; \mathcal{K})}{J_N(h;\mathcal{O})}
    & =
    \mathrm{e}^{- \frac{1}{2} m h (N^2-1) - \frac{m^2+1}{4 m} h}
    \sum_{\varepsilon=\pm 1}
    \sum_{j=0}^{N-1}
    \varepsilon \,
    \mathrm{e}^{m h (j+\frac{m+\varepsilon}{2 m})^2}
    \\
    & =
    \frac{1}{\sqrt{\pi \, m \, h}} \,
    \mathrm{e}^{-\frac{1}{2} m h (N^2-1) - \frac{m^2+1}{4 m} h}
    \sum_{\varepsilon=\pm 1}
    \sum_{j=0}^{N-1}
    \varepsilon \,
    \int\limits_\mathcal{C} \mathrm{d} z \,
    \mathrm{e}^{-\frac{z^2}{m h} + 2 ( j+\frac{m+\varepsilon}{2 m}) z}
    \\
    & =
    \frac{1}{\sqrt{\pi \, m \, h}} \,
    \mathrm{e}^{- \frac{1}{2} m h N^2 +\frac{m}{4} h - \frac{h}{4 m}}
    \, 
    \int\limits_\mathcal{C} \mathrm{d} z \,
    \mathrm{e}^{- \frac{z^2}{m h} + N z} \,
    \frac{
      2 \sh(N \, z) \, \sh(z/m)
    }{
      \sh(z)
    } .
  \end{align*}
  Now we take a limit
  \begin{equation*}
    h \to \frac{2 \, \pi \, \mathrm{i}}{N} ,
  \end{equation*}
  to compute Kashaev's invariant.
  As LHS vanishes in this limit,
  we take a derivative of both  sides to have
  \begin{equation*}
    \langle T(2,2 \, m) \rangle_N
    =
    \frac{1}{4 \,  \sqrt{2} \,  \pi^3}
    \left( \frac{\   N  \   }{m} \right)^{3/2}  \,
    \mathrm{e}^{\frac{\pi \mathrm{i}}{2 N}(m - \frac{1}{m})
      - \frac{\pi \mathrm{i}}{4}} \,
    \int\limits_\mathcal{C} \mathrm{d} z \,
    z^2 \,
    \mathrm{e}^{N (\frac{\mathrm{i} z^2}{2 m \pi}+z)} \,
    \frac{
      2 \sh(N \, z) \, \sh(z/m)
    }{
      \sh(z)
    } .
  \end{equation*}
  Rescaling an integral variable, we obtain eq.~\eqref{link_integral}.  
\end{proof}

\begin{theorem}
  \label{theorem:asymptotic}
  Kashaev's invariant $\langle \mathcal{K} \rangle_N$ for the torus
  link $\mathcal{K}=T(2,2 \, m)$ has an asymptotic expansion in
  $N\to\infty$ as
  \begin{multline}
    \langle T(2, 2 \, m) \rangle_N
    \simeq
    \mathrm{e}^{\frac{3}{4} \pi \mathrm{i}
      -
      \frac{(m-1)^2}{2 m N}  \pi \mathrm{i}
    } \,
    N^{3/2} \,
    \sqrt{\frac{\   2  \  }{m}}
    \sum_{k=1}^{m-1}
    (-1)^k \, (k-m) \,
    \sin
    \left(\frac{k}{\   m  \  } \,  \pi \right)
    \,
    \mathrm{e}^{-\frac{k^2}{2 m } \pi \mathrm{i} N}
    \\
    +
    \mathrm{e}^{
      -
      \frac{(m-1)^2}{2 m N}  \pi \mathrm{i}
    } \,
    N \,
    \sum_{k=0}^\infty
    \frac{
      E_{ k}^{(m; 0)}
      }{k!} \,
    \left(
      \frac{\pi \, \mathrm{i}}{2 \, m \, N}
    \right)^k    ,
  \end{multline}
  where $E_k^{(m; 0)}$ is defined from a generating function
  (see also eq.~\eqref{E_and_Bernoulli})
  as
  \begin{align}
    \label{define_euler}
    \frac{m \sh(z)}{\sh ( m \, z)}
    & =
    \sum_{k=0}^\infty
    \frac{E_{k}^{(m; 0)}}{(2 \, k)!} \, z^{2 k} 
    \\
    \nonumber
    & =
    1 + \frac{1-m^2}{6} \, z^2 
    +
    \frac{(3 - 7 \, m^2 ) \, (1- m^2)}{360} \, z^4 + \cdots .
  \end{align}
\end{theorem}

\begin{proof}
  When we decompose
  $\sh(4 \, m \, \pi \, N\, w)$
  in the integrand~\eqref{link_integral}
  into
  $(\mathrm{e}^{4 m \pi N w} - \mathrm{e}^{- 4 m \pi N w})/2$,
  the integral reduces to $I_1 - I_2$ up to constant where $I_1$ and
  $I_2$ are 
  \begin{align*}
    I_1
    & =
    \int\limits_\mathcal{C} \mathrm{d} w \, w^2 \,
    \mathrm{e}^{8 m \pi N (\mathrm{i} w^2 + w)} \,
    \frac{
      \sh(4 \, \pi \, w)}
    {\sh (4 \, m  \, \pi \, w)} ,
    \\[2mm]
    I_2
    & =
    \int\limits_{\mathcal{C}} \mathrm{d} w \,
    w^2 \,
    \mathrm{e}^{8 m \pi N \mathrm{i} w^2}
    \frac{\sh(4 \, \pi \, w)}
    {\sh(4 \, m \, \pi \, w)} .
  \end{align*}
  In $I_1$ we deform an integration path
  $\mathcal{C} \to \mathcal{C} + \frac{\mathrm{i}}{2}$.
  In this deformation  we have contributions from
  residues at
  $w=\frac{k}{4 \, m} \, \mathrm{i}$
  for $k=1,2, \dots,  2 \, m -1$.
  Then we get
  \begin{align*}
    I_1
    & =
    2 \, \pi \, \mathrm{i} \cdot
    \left(
      \text{Residues}
    \right)
    +
    \int\limits_{
      \mathcal{C}+\frac{\mathrm{i}}{2}
    }
    \mathrm{d} w \,
    w^2 \,
    \mathrm{e}^{8 m \pi N (\mathrm{i} w^2+w)}
    \, \frac{\sh(4 \, \pi \, w)}{\sh ( 4 \, m \, \pi \, w)}
    \\
    & =
    \frac{1}{32 \, m^3}
    \sum_{k=1}^{2 \, m-1}
    (-1)^k \, k^2
    \sin
    \left(
      \frac{k}{m} \,\pi
    \right) \,
    \mathrm{e}^{-\frac{k^2}{2 m} \pi \mathrm{i} N}
    +
    \int\limits_\mathcal{C} \mathrm{d} z \,
    \left(
      z+ \frac{\mathrm{i}}{2}
    \right)^2 \,
    \mathrm{e}^{8 m \pi \mathrm{i} N  z^2}
    \, \frac{\sh(4 \, \pi \, z)}{\sh ( 4 \, m \, \pi \, z)}
    \\
    & =
    \frac{1}{8 \, m^2}
    \sum_{k=1}^{m-1}
    (-1)^k \, (k-m) \,
    \sin
    \left(
      \frac{k}{m} \, \pi
    \right) \, \mathrm{e}^{- \frac{k^2}{2 m} \pi \mathrm{i} N}
    + I_2
    -
    \frac{1}{4} \int\limits_\mathcal{C} \mathrm{d} z \,
    \mathrm{e}^{8 m \pi \mathrm{i} N z^2}
    \,
    \frac{\sh(4 \, \pi \, z)}{\sh(4 \, m \, \pi \, z)} .
  \end{align*}
  In the last equality, we have used a symmetry
  $z\leftrightarrow  -z$ of the integrand.
  Substituting
  a series expansion~\eqref{define_euler}  into
  above expression, we obtain an assertion of theorem.
\end{proof}

Asymptotic expansion for the torus knot $T(2,2\,m+1)$ was studied in
Ref.~\citen{KHikami02c}.
In view of these results,
a tail of asymptotic
expansion of Kashaev's invariant
is given in an infinite series of $N^{-1}$ with coefficients,
whose generating function
seems to be related to
\begin{equation*}
  \frac{1-s}{\Delta_\mathcal{K}(s)} .
\end{equation*}
Here
$\Delta_\mathcal{K}(s)$
is the Alexander polynomial for knot $\mathcal{K}$,
and 
in a  case of the torus link $\mathcal{K}=T(2,2\,m)$ we have
\begin{equation*}
  \Delta_\mathcal{K}(s)=
  \frac{1-s^{2m}}{1+s} .
\end{equation*}

We have shown   that the volume conjecture~\cite{Kasha95,MuraMura99a}
is fulfilled for the torus
link $T(2,2\,m)$,
\begin{equation*}
  \lim_{N\to\infty}
  \frac{2 \, \pi }{N} \,
  \log
  \bigl|
    \langle T(2, 2\,m) \rangle_N
  \bigr|
  =0 ,
\end{equation*}
because the torus link $T(2,2 \, m)$ is not hyperbolic.
Rather we have interests in an asymptotic expansion of the $q$-series.
To this aim, we  compute the quantum invariant by another method;
explicit form of Kashaev's invariants can be directly
computed from a set of the enhanced Yang--Baxter operators~\cite{Kasha95}
(see also Refs.~\citen{YYokot00b,MuraMura99a,HMuraka00c});
\begin{subequations}
  \label{R_matrix}
  \begin{gather}
    R^{i j}_{k \ell}
    =
    \frac{N \, \omega^{1-(k-j+1)(\ell-i)} 
    }{
      (\omega)_{[\ell-k-1]} \, (\omega)_{[j-\ell]}^{~*} \,
      (\omega)_{[i-j]} \, (\omega)_{[k-i]}^{~*}
    } \cdot
    \theta
    \begin{bmatrix}
      i & j      \\
      k & \ell
    \end{bmatrix} ,
    \\[2mm]
    (R^{-1})^{i j}_{k \ell}
    =
    \frac{
      N \, \omega^{-1+(\ell-i-1)(k-j)}
    }{
      (\omega)_{[\ell-k-1]}^{~*} \,
      (\omega)_{[j-\ell]} \, (\omega)_{[i-j]}^{~*} \,
      (\omega)_{[k-i]}
    } \cdot
    \theta
    \begin{bmatrix}
      i & j \\
      k & \ell
    \end{bmatrix}  ,
    \\[2mm]
    \mu^k_\ell
    = - \delta_{k, \ell+1} \omega^{1/2} ,
  \end{gather}
\end{subequations}
where
we have defined the $N$-th root of unity as
\begin{equation}
  \label{q_prod}
  \omega = \exp \left(
    \frac{2 \, \pi \, \mathrm{i}}{N}
  \right) ,
\end{equation}
and
$*$ means a complex conjugation.
We have  also used
$[x] \in \{0,1,\dots,N-1\}$ modulo $N$, and
\begin{equation*}
  \theta 
  \begin{bmatrix}
    i & j \\
    k & \ell
  \end{bmatrix}
  =1 ,\quad
  \text{if and only if}~
  \begin{cases}
    i \leq k < \ell \leq  j,
    \\
    j \leq i \leq k < \ell ,
    \\
    \ell \leq j \leq i \leq k ~(\text{with $\ell<k$}),
    \\
    k< \ell \leq j \leq i  .
  \end{cases}
\end{equation*}
Remark that used is  the standard notation of the $q$-product and the
$q$-binomial coefficient;
\begin{gather}
  (\omega)_n = \prod_{k=1}^n ( 1 - \omega^k )  ,
  \\[2mm]
  \label{q_binomial}
  \begin{bmatrix}
    m \\ n
  \end{bmatrix}
  =
  \begin{cases}
    \displaystyle
    \frac{(\omega)_m}{(\omega)_n \, (\omega)_{m-n}} ,
    &
    \text{if $m\geq n \geq 0$,}
    \\[3mm]
    0 ,
    &
    \text{others.}
  \end{cases}
\end{gather}
These operators are assigned to a projection of knot as
follows;
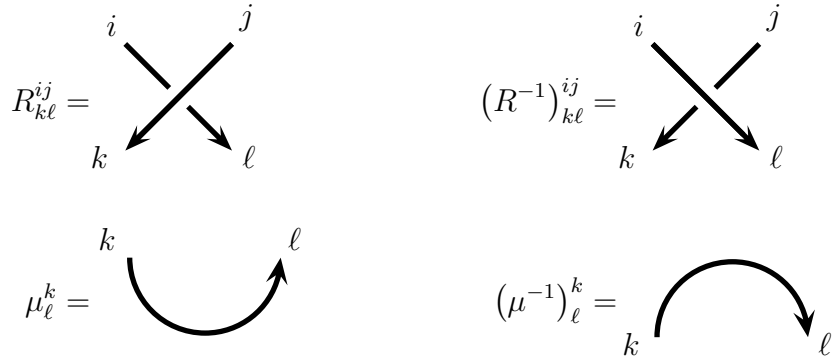
\begin{align*}
  R^{i j}_{k \ell}
  &=
  \begin{psfrags}
    \psfrag{i}{$i$}
    \psfrag{j}{$j$}
    \psfrag{k}{$k$}
    \psfrag{l}{$\ell$}
    \includegraphics[scale=1, bb=-30 -5 10 35]{R.ps}
  \end{psfrags}
  &
  \bigl( R^{-1} \bigr)^{i j}_{k \ell}
  & =
  \begin{psfrags}
    \psfrag{i}{$i$}
    \psfrag{j}{$j$}
    \psfrag{k}{$k$}
    \psfrag{l}{$\ell$}
    \includegraphics[scale=1, bb=-30 -5 10 35]{Rinv.ps}
  \end{psfrags}
  \\[4mm]
  \mu^k_\ell
  & =
  \begin{psfrags}
    \psfrag{k}{$k$}
    \psfrag{l}{$\ell$}
    \includegraphics[scale=1, bb=-40 -20 10 35]{mu.ps}
  \end{psfrags}
  &
  \bigl(\mu^{-1}\bigr)^k_\ell
  &=
  \begin{psfrags}
    \psfrag{k}{$k$}
    \psfrag{l}{$\ell$}
    \includegraphics[scale=1, bb=-40 10 10 35]{muinv.ps}
  \end{psfrags}
  \\[3mm]
\end{align*}

\begin{prop}
  \label{prop:explicit}
  Kashaev's invariant $\langle \mathcal{K} \rangle_N$ for the torus
  link $\mathcal{K}=T(2,2 \, m)$ is explicitly given by
  \begin{multline}
    \langle T(2, 2 \, m) \rangle_N
    \\
    =
    N \,
    \sum_{N-1 \geq c_{m-1} \geq  \dots \geq c_2 \geq c_1 \geq 0}
    (-1)^{c_{m-1}} \,
    \omega^{\frac{1}{2} c_{m-1} (c_{m-1}+1)}
    \,
    \left(
      \prod_{i=1}^{m-2}
      \omega^{c_i (c_i + 1)} \,
      \begin{bmatrix}
        c_{i+1}
        \\
        c_i
      \end{bmatrix}
    \right) .
  \end{multline}
\end{prop}

\begin{proof}
  We get this result  from a direct computation using
  $R$-matrix~\eqref{R_matrix} for ($1,1$)-tangle of ($2,2 \, m$)-torus
  link.
  See Refs.~\citen{YYokot00b,HMuraka00c}.
\end{proof}

We note that Kashaev's invariant for the Hopf link $T(2,2)$ is given
by
\begin{equation*}
  \langle T(2,2) \rangle_N
  = N .
\end{equation*}

Combining Prop.~\ref{prop:explicit} with
Thm.~\ref{theorem:asymptotic},
we obtain an asymptotic expansion of $\omega$-series.
\begin{coro}
  \label{Coro:asymptotic}
  \begin{multline}
    \label{torus_link_asymptotic}
    \sum_{N-1 \geq c_{m-1} \geq  \dots \geq c_2 \geq c_1 \geq 0}
    (-1)^{c_{m-1}} \,
    \omega^{\frac{1}{2} c_{m-1} (c_{m-1}+1)}
    \,
    \left(
      \prod_{i=1}^{m-2}
      \omega^{c_i (c_i + 1)} \,
      \begin{bmatrix}
        c_{i+1}
        \\
        c_i
      \end{bmatrix}
    \right)
    \\
    \simeq
    \sqrt{N} \,
    \mathrm{e}^{\frac{3}{4} \pi \mathrm{i}
      - \frac{ (m-1)^2}{2 m N} \pi \mathrm{i}} \,
    \sqrt{\frac{2}{  \   m  \  }} \, 
    \sum_{k=1}^{m-1}
    (-1)^k \,
    (k-m) \sin \left( \frac{k}{ \  m  \  } \, \pi \right) \,
    \mathrm{e}^{- \frac{k^2}{2 m} \pi \mathrm{i} N}
    \\
    +
    \mathrm{e}^{
      - \frac{ (m-1)^2}{2 m N}  \pi \mathrm{i}} \,
     \sum_{k=0}^\infty
    \frac{E_{ k}^{(m; 0)}}{k!} \,
    \left(
      \frac{ \pi \, \mathrm{i}}{2 \, m \, N}
    \right)^k .
  \end{multline}
\end{coro}


In the rest of this paper, we shall reveal a meaning of this
asymptotic expansion from a point of view of the modular form.
As a generalization of $\omega$-series defined by Kashaev's invariant
we introduce
\begin{multline}
  \label{define_Y_a}
  Y_m^{(a)}(\omega)
  =
  \sum_{c_1, \dots , c_{m-1}=0}^{N-1}
  (-1)^{c_{m-1}} \,
  \omega^{\frac{1}{2} c_{m-1} (c_{m-1}+1)}
  \\
  \times
  \omega^{c_1^{~2} + \dots + c_{m-2}^{~2}+c_{a+1}+\dots +c_{m-2}}
  \left(
    \prod_{i=1}^{m-2}
    \begin{bmatrix}
      c_{i+1} + \delta_{i,a} \\
      c_i
    \end{bmatrix}
  \right) ,
\end{multline}
for $m\geq 2$ and $a=0, 1, \dots, m-2$.
See that Kashaev's invariant for $T(2, 2 \, m)$ corresponds to a case of
$a=0$,
\begin{equation}
  \label{Invariant_and_Y}
  \langle T(2, 2 \, m) \rangle_N
  =
  N \cdot Y_m^{(0)}(\omega) .
\end{equation}
It is unclear whether
the $\omega$-series $Y_m^{(a)}(\omega)$ for $a\neq 0$ represent the
quantum invariant for any  three manifolds.

\begin{conj}
  \label{conj:1}
  Let $Y_m^{(a)}(\omega)$ be defined by eq.~\eqref{define_Y_a}.
  An asymptotic expansion of this $\omega$-series
  in $N\to \infty$ is given by
  \begin{multline}
    \mathrm{e}^{
      \frac{\pi \mathrm{i}}{2 m N} (m-1-a)^2
    }
    \cdot
    Y_m^{(a)}(\omega)
    \\
    \simeq
    \sqrt{N} \,
    \mathrm{e}^{
      \frac{3}{4} \pi \mathrm{i}
    }
    \cdot
    \sqrt{\frac{2}{  \  m   \  }} \,
    \sum_{k=1}^{m-1}
    (-1)^k \, (k-m) \,
    \sin
    \left(
      \frac{k}{  \  m  \  } \, (a+1) \, \pi
    \right) \,
    \mathrm{e}^{- \frac{k^2}{2 m} \pi \mathrm{i} N}
    \\
    +
    \sum_{k=0}^\infty
    \frac{E_{k}^{(m ; a)}}{k !}
    \left(
      \frac{\pi \, \mathrm{i}}{2 \, m \, N}
    \right)^k ,
    \label{conjecture_1}
  \end{multline}
  where a generalized Euler number $E_k^{(m;a)}$ is given from a
  generating function as
  \begin{align}
    \label{generate_E_n}
    \frac{
      m \, \sh \bigl( (a+1) \, z \bigr)
    }{
      \sh ( m \, z)}
    & =
    \sum_{k=0}^{\infty}
    \frac{E_{k}^{(m ; a)}}{(2 \, k )!} \, z^{2 k}
    \\
    \nonumber
    & =
    (a+1) + \frac{1}{6} \,(1+a) \,
    \bigl( (1+a)^2 -m^2 \bigr) \, z^2
    \\
    \nonumber
    & \qquad  
    +
    \frac{1}{360} \,
    (1+a) \, \bigl( (1+a)^2 - m^2\bigr) \,
    \bigl( 3 \, (1+a)^2 - 7 \, m^2 \bigr) \, z^4
    \\
    \nonumber
    & 
    +
    \frac{1}{15120} \,
    (1+a) \, \bigl( (1+a)^2 - m^2 \bigr) \,
    \bigl(
    3 \, (1+a)^4 - 18 \, (1+a)^2 \, m^2 + 31 \, m^4
    \bigr) \, z^6
    + \cdots .
  \end{align}
\end{conj}

The  case  $a=0$ of this conjecture
is proved in Corollary~\ref{Coro:asymptotic}.

Note that we have
\begin{equation}
  \label{generate_chi}
  \frac{ \sh\bigl((a+1)z \bigr)}{\sh ( m \, z)}
  =
  \sum_{n=0}^\infty
  \chi_{2 m}^{(a)}(n) \,
  \mathrm{e}^{- n z} ,
\end{equation}
where the odd periodic function $\chi_{2 m}^{(a)}(n)$ is written as  
\begin{equation}
  \label{define_chi}
  \begin{array}{c|ccc}
    n \mod 2 \, m & m-1-a & m+1+a & \text{others}
    \\
    \hline
    \chi_{2 m}^{(a)}(n)
    & 1 & -1 & 0
  \end{array}
\end{equation}
Applying the Mellin transformation to eqs.~\eqref{generate_E_n}
and~\eqref{generate_chi}, we have an expression of the generalized
Euler number in terms of the $L$-function associated to
$\chi_{2 m}^{(a)}(n)$;
\begin{align}
  E_k^{(m;a)}
  & = m \cdot L(-2 \, k , \chi_{2 m}^{(a)})
  \nonumber 
  \\
  & =
  - m \,
  \frac{ ( 2 \, m)^{2 k}}{2 \, k+1} \,
  \left(
    B_{2 k+1}
    \left(
      \frac{m-1-a}{2 \, m}
    \right)
    -
    B_{2 k+1}
    \left(
      \frac{m+1+a}{2 \, m}
    \right)
  \right) ,
  \label{E_and_Bernoulli}
\end{align}
where $B_n(x)$ is the $n$-th Bernoulli polynomial.
It should be remarked that the colored Jones polynomial~\eqref{Jones_link}
for the torus link $\mathcal{K}=T(2, 2 \, m)$ is rewritten using the
periodic function as
\begin{equation}
  2 \sh \left(\frac{N \, h}{2}\right) \,
  \frac{
    J_N(h;\mathcal{K})
  }{
    J_N(h;\mathcal{O})
  }
  =
  -\mathrm{e}^{
    -\frac{1}{2} m (N^2 -1) h - \frac{m^2+1}{4 m} h
  }
  \sum_{k=0}^{2 m N}
  \chi_{2 m}^{(0)}(k) \,
  \mathrm{e}^{\frac{k^2}{4 m} h} .
\end{equation}
Based on this expression, we find that Kashaev's invariant is given by
\begin{equation}
  \label{Other_Expression_Invariant}
  \langle T(2, 2 \, m) \rangle_N
  =
  - \frac{1}{4 \, m \, N} \,
  \mathrm{e}^{-\frac{(m-1)^2}{2 m N} \pi \mathrm{i}} \,
  \sum_{k=0}^{2 m N}
  k^2 \, \chi_{2 m}^{(0)}(k) \,
  \mathrm{e}^{\frac{k^2}{2 m N} \pi \mathrm{i}} .
\end{equation}

Later we shall clarify a relationship between above conjecture and the
modular form.

\section{\mathversion{bold}
  $q$-Series Identity}
\label{sec:q_series}

In this section we study  a $q$-series identity, which is closely
related with $Y_m^{(a)}(\omega)$ defined in eq.~\eqref{define_Y_a}.
We use standard notation as in
eqs.~\eqref{q_prod}~--~\eqref{q_binomial}, but in this section
we replace $\omega$, the $N$-th primitive root of unity,
with generic $q$.

We define the $q$-series
\begin{multline}
  \label{define_K_m}
  K_m^{(a)}(x)
  =
  \sum_{c_1, \dots, c_{m-1}=0}^\infty
  (-1)^{c_{m-1}} \, 
  q^{\frac{1}{2} c_{m-1} (c_{m-1} +1)} \,
  x^{c_1 + \dots + c_{m-1}}
  \\
  \times
  q^{c_1^{~2} + \dots + c_{m-2}^{~2} + c_{a+1} + \dots + c_{m-2}}
  \cdot
  \left(
    \prod_{i=1}^{m-2}
    \begin{bmatrix}
      c_{i+1} + \delta_{i,a} \\
      c_i
    \end{bmatrix}
  \right) ,
\end{multline}
for $m\geq 2$ and  $a=0,1, 2,\dots, m-2$.
We simply replace $\sum_{c_{m-1}=0}^{N-1}$ in eq.~\eqref{define_Y_a}
with an infinite sum
$\sum_{c_{m-1}=0}^{\infty}$,
though we have  introduced an additional variable $x$.

\begin{theorem}
  Let the $q$-series $K_m^{(a)}(x)$ be defined in
  eq.~\eqref{define_K_m}.
  We have
  \begin{equation}
    \label{q_series_main}
    K_m^{(a)}(x)
    =
    \sum_{n=0}^\infty
    \chi_{2 m}^{(a)}(n) \,
    q^{\frac{n^2 - (m-1-a)^2}{4m}} \,
    x^{\frac{n-(m-1-a)}{2}} ,
  \end{equation}
  where $\chi_{2 m}^{(a)}(n)$ is a periodic function
  in eq.~\eqref{define_chi}.
\end{theorem}

\begin{proof}
  We prove this statement  by showing that
  both sides satisfy the same
  $q$-difference equation.

  It is easy to see from a periodicity of the function
  $\chi_{2 m}^{(a)}(n)$
  that RHS of eq.~\eqref{q_series_main} solves a difference
  equation
  (see, \emph{e.g.}, Refs.~\citen{Andre76,DZagie01a,KHikami02c}),
  \begin{align}
    K_m^{(a)}(x)
    & =
    1- q^{a+1} \, x^{a+1}
    +
    \sum_{n=2m}^\infty
    \chi_{2 m}^{(a)}(n) \,
    q^{\frac{n^2 - (m-1-a)^2}{4m}} \,
    x^{\frac{n-(m-1-a)}{2}} 
    \nonumber \\
    & =
    1- q^{a+1} \, x^{a+1} 
    +
    x^m \, q^{2 m -1-a} \,
    K_m^{(a)}(q^2 \, x) .
    \label{difference_K}
  \end{align}
  We shall show that LHS of eq.~\eqref{q_series_main}
  also fulfills a same difference equation.
  To this aim, we introduce
  \begin{multline}
    \label{define_K_m_multi_x}
    K_m^{(a)}(x_1, \dots, x_{m-1})
    =
    \sum_{c_1 , \dots, c_{m-1}=0}^\infty
    (-1)^{c_{m-1}} \,
    q^{\frac{1}{2} c_{m-1} (c_{m-1} + 1)} \, x_{m-1}^{~c_{m-1}}
    \\
    \times
    \left(
      \prod_{i=1}^{a-1} q^{c_i^{~2}} \, x_i^{~c_i} \,
      \begin{bmatrix}
        c_{i+1} \\
        c_i
      \end{bmatrix}
    \right)
    \cdot
    q^{c_a^{~2}} \, x_a^{~c_a} \,
    \begin{bmatrix}
      c_{a+1} + 1 \\
      c_a
    \end{bmatrix}
    \\
    \times
    \left(
      \prod_{i=a+1}^{m-2}
      q^{c_i^{~2} + c_i} \, x_i^{~c_i} \,
      \begin{bmatrix}
        c_{i+1} \\
        c_i
      \end{bmatrix}
    \right) .
  \end{multline}
  See that by definition we have
  \begin{equation}
    K_m^{(a)}(x)
    =
    K_m^{(a)}(
    \underbrace{x, \dots, x}_{m-1}) .
  \end{equation}
  We use same symbol $K_m^{(a)}$, but we believe there is no confusion.

  To prove the assertion of the theorem,
  we use formulae for the $q$-binomial coefficients;
  \begin{subequations}
    \begin{align}
      \begin{bmatrix}
        n+1 \\
        c
      \end{bmatrix}
      & =
      q^c \,
      \begin{bmatrix}
        n \\
        c
      \end{bmatrix}
      +
      \begin{bmatrix}
        n \\
        c-1
      \end{bmatrix}
      \label{binomial_1}
      \\
      & =
      \begin{bmatrix}
        n \\
        c
      \end{bmatrix}
      +
      q^{n+1-c} \,
      \begin{bmatrix}
        n \\
        c-1
      \end{bmatrix} .
      \label{binomial_2}
    \end{align} 
  \end{subequations}
  Applying eq.~\eqref{binomial_1} to
  $
  \begin{bmatrix}
    c_{a+1}+1 \\
    c_a
  \end{bmatrix}
  $ in eq.~\eqref{define_K_m_multi_x}, we get
  \begin{multline}
    \label{maru_1}
    K_m^{(a)}(x_1, \dots, x_{m-1})
    =
    K_m^{(0)}(q^{-1} \, x_1, \dots, q^{-1} \, x_{a-1} ,
    x_a, \dots, x_{m-1})
    \\
    +
    q \, x_a \cdot
    K_m^{(a-1)}(x_1, \dots, x_{a-1} , q \, x_a ,
    x_{a+1} , \dots, x_{m-1}) .
  \end{multline}
  On the other hand, using eq.~\eqref{binomial_2}, we have
  \begin{multline}
    \label{maru_2}
    K_m^{(a)}(x_1, \dots, x_{m-1})
    =
    K_m^{(0)}(q^{-1} \, x_1, \dots, q^{-1} \, x_{a} ,
    x_{a+1}, \dots, x_{m-1})
    \\
    +
    q \, x_a \cdot
    K_m^{(a-1)}(x_1, \dots, x_{a} , q \, x_{a+1} ,
    x_{a+2} , \dots, x_{m-1})
  \end{multline}
  Another difference equation is given as follows;
  \begin{align}
    & K_m^{(0)}(x_1, \dots, x_{m-1})
    \nonumber \\
    & =
    1+
    \sum_{c_{m-1}=1}^\infty
    \sum_{c_1, \dots, c_{m-2}=0}^\infty
    (-1)^{c_{m-1}} \, q^{\frac{1}{2} c_{m-1} (c_{m-1} +1)} \,
    x_{m-1}^{~c_{m-1}}
    \,
    \left(
      \prod_{i=1}^{m-2}
      q^{c_i (c_i+1)} \, 
      x_i^{~c_i} \,
      \begin{bmatrix}
        c_{i+1} \\
        c_i
      \end{bmatrix}
    \right)
    \nonumber \\
    & =
    1 - q \, x_{m-1} \cdot
    K_m^{(m-2)}(q \, x_1, \dots, q \, x_{m-1}) .
    \label{maru_3}
  \end{align}

  We can prove eq.~\eqref{difference_K} by use of
  eqs.~\eqref{maru_1}---\eqref{maru_3} as follows.
  Recursive use of eq.~\eqref{maru_1} gives
  \begin{multline}
    \label{a_and_0_1}
    K_m^{(a)}(x_1, \dots,x_a ,
    q\, x_{a+1},\dots,q \, x_{m-1})
    \\
    =
    K_m^{(0)}(q^{-1} \, x_1 , \dots, q^{-1} \, x_{a-1},
    x_a , q \, x_{a+1} , \dots, q \, x_{m-1})
    \\
    + q \, x_a \cdot
    K_m^{(0)}(q^{-1} \, x_1 , \dots, q^{-1} \, x_{a-2},
    x_{a-1} , q \, x_{a} , \dots, q \, x_{m-1})
    \\
    + q^2 \, x_{a-1} \, x_a
    \cdot
    K_m^{(0)}(q^{-1} \, x_1 , \dots, q^{-1} \, x_{a-3},
    x_{a-2} , q \, x_{a-1} , \dots, q \, x_{m-1})
    \\
    + \dots
    +
    q^a \, x_1 \cdots x_a \cdot
    K_m^{(0)}(q \, x_1 , \dots, q \, x_{m-1}) .
  \end{multline}
  Substituting above equation for $a=m-2$ into eq.~\eqref{maru_3}, we
  get
  \begin{multline}
    \label{comp_1}
    K_m^{(0)}(q^{-1} \, x_1, \dots , q^{-1} x_{m-2}, x_{m-1})
    \\
    +
    q \, x_{m-1} \cdot
    K_m^{(0)}(q^{-1} \, x_1, \dots, q^{-1} x_{m-3},
    x_{m-2}, q \, x_{m-1})
    \\
    +
    q^2 \, x_{m-2} \, x_{m-1} \cdot
    K_m^{(0)}(q^{-1} \, x_1, \dots, q^{-1} x_{m-4},
    x_{m-3}, q \, x_{m-2}, q \, x_{m-1})
    \\
    + \dots
    +
    q^{m-2} \, x_{2} \cdots x_{m-1} \cdot
    K_m^{(0)}(x_1, q\, x_2 , \dots, q \, x_{m-1})
    \\
    =
    1 - q^{m-1} \, x_1 \cdots x_{m-1}
    \cdot
    K_m^{(0)}(q \, x_1, \dots, q \, x_{m-1}) .
  \end{multline}

  In the same way,
  iterated use of eq.~\eqref{maru_2} gives
  \begin{multline}
    \label{a_and_0_2}
    K_m^{(a)}(x_1, \dots,x_{a+1} ,
    q\, x_{a+2},\dots,q \, x_{m-1})
    \\
    =
    K_m^{(0)}(q^{-1} \, x_1 , \dots, q^{-1} \, x_{a},
    x_{a+1} , q \, x_{a+2} , \dots, q \, x_{m-1})
    \\
    + q \, x_a \cdot
    K_m^{(0)}(q^{-1} \, x_1 , \dots, q^{-1} \, x_{a-1},
    x_{a} , q \, x_{a+1} , \dots, q \, x_{m-1})
    \\
    + q^2 \, x_{a-1} \, x_a
    \cdot
    K_m^{(0)}(q^{-1} \, x_1 , \dots, q^{-1} \, x_{a-2},
    x_{a-1} , q \, x_{a} , \dots, q \, x_{m-1})
    \\
    + \dots
    +
    q^a \, x_1 \cdots x_a \cdot
    K_m^{(0)}( x_1 , q \, x_2,  \dots, q \, x_{m-1}) .
  \end{multline}
  Substituting above equation for $a=m-2$ into eq.~\eqref{maru_3}, we
  find
  \begin{multline}
    \label{comp_2}
    K_m^{(0)}(q^{-1} \, x_1, \dots , q^{-1} x_{m-2}, x_{m-1})
    \\
    +
    q \, x_{m-2} \cdot
    K_m^{(0)}(q^{-1} \, x_1, \dots, q^{-1} x_{m-3},
    x_{m-2}, q \, x_{m-1})
    \\
    +
    q^2 \, x_{m-3} \, x_{m-2} \cdot
    K_m^{(0)}(q^{-1} \, x_1, \dots, q^{-1} x_{m-4},
    x_{m-3}, q \, x_{m-2}, q \, x_{m-1})
    \\
    + \dots
    +
    q^{m-2} \, x_{1} \cdots x_{m-2} \cdot
    K_m^{(0)}(x_1, q\, x_2 , \dots, q \, x_{m-1})
    \\
    =
    \frac{1}{x_{m-1}} \,
    \left(
      1 - 
      K_m^{(0)}(q^{-1} \, x_1, \dots, q^{-1} \, x_{m-1})
    \right) .
  \end{multline}
  Combining  two equations~\eqref{comp_1} and~\eqref{comp_2} with
  $x \equiv x_1 = \dots = x_{m-1}$, we obtain
  \begin{equation*}
    K_m^{(0)}(x)
    =
    1 - q \, x
    + q^{2 m -1} \, x^m \cdot
    K_m^{(0)}(q^2 \, x) .
  \end{equation*}
  This proves eq.~\eqref{q_series_main} for $a=0$.

  Other cases can be shown
  by using eqs.~\eqref{comp_1} and~\eqref{comp_2} with
  eqs.~\eqref{a_and_0_1} and~\eqref{a_and_0_2}.
\end{proof}

\begin{coro}
  We define the $q$-series
  $\widetilde{\Phi}_m^{(a)}(\tau)$ by
  $K_m^{(a)}(x=1)$ up to constant,
  \emph{i.e.},
  \begin{multline}
    \label{define_Phi_m}
    \widetilde{\Phi}_m^{(a)}(\tau)
    =
    m \,
    q^{\frac{(m-1-a)^2}{4 m}} \,
    \sum_{c_1, \dots, c_{m-1}=0}^\infty
    (-1)^{c_{m-1}} \, 
    q^{\frac{1}{2} c_{m-1} (c_{m-1} +1)} 
    \\
    \times
    q^{c_1^{~2} + \dots + c_{m-2}^{~2} + c_{a+1} + \dots + c_{m-2}}
    \cdot
    \left(
      \prod_{i=1}^{m-2}
      \begin{bmatrix}
        c_{i+1} + \delta_{i,a} \\
        c_i
      \end{bmatrix}
    \right) ,
  \end{multline}
  where  $m \geq 2$ and $a=0,1,\dots,m-2$,
  and we mean
  \begin{equation*}
    q=\mathrm{e}^{2 \pi \mathrm{i} \tau} .
  \end{equation*}
  Then we have
  \begin{equation}
    \label{Eichler_Phi}
    \widetilde{\Phi}_m^{(a)}(\tau)
    =
    m \, \sum_{n=0}^\infty
    \chi_{2 m}^{(a)}(n) \,
    q^{\frac{1}{4m}  n^2} .
  \end{equation}
\end{coro}

Factor $m$ in above definition is merely for our later convention.

\begin{coro}
  Let the $q$-series $\Phi_m^{(a)}(\tau)$ be defined by
  \begin{equation}
    \label{define_modular}
    \Phi_m^{(a)}(\tau)
    =
    \sum_{n \in \mathbb{Z}}
    n \, \chi_{2 m}^{(a)}(n) \,
    q^{\frac{1}{4 m} n^2 } ,
  \end{equation}
  for $m \geq 2$ and $a=0,1,\dots, m-2$.
  Then we have
  \begin{multline}
    \label{q_series_modular}
    \Phi_m^{(a)}(\tau)
    \\
    = 4 \, q^{\frac{(m-1-a)^2}{4 m}} \,
    \sum_{c_1,\dots,c_{m-1}=0}^\infty
    \left(
      c_1 + c_2 + \dots + c_{m-1} + \frac{m-1-a}{2}
    \right) \,
        (-1)^{c_{m-1}} \, 
    q^{\frac{1}{2} c_{m-1} (c_{m-1} +1)} 
    \\
    \times
    q^{c_1^{~2} + \dots + c_{m-2}^{~2} + c_{a+1} + \dots + c_{m-2}}
    \cdot
    \left(
      \prod_{i=1}^{m-2}
      \begin{bmatrix}
        c_{i+1} + \delta_{i,a} \\
        c_i
      \end{bmatrix}
    \right) .
  \end{multline}
\end{coro}

\begin{proof}
  We differentiate eq.~\eqref{q_series_main} with respect to $x$ and
  substitute $x\to 1$.
\end{proof}

\section{Modular Property}
\label{sec:modular}

We shall reveal  the modular property of the $q$-series
$\Phi_m^{(a)}(\tau)$ and
$\widetilde{\Phi}_m^{(a)}(\tau)$
defined by eq.~\eqref{define_modular} and eq.~\eqref{Eichler_Phi}
respectively.
The  theta series $\Phi_m^{(a)}(\tau)$
have weight $3/2$, and
span ($m-1$)-dimensional space;
it is straightforward to get
\begin{gather}
  \Phi_m^{(m-1-a)}(\tau+1)
  =
  \mathrm{e}^{\frac{a^2}{2 m} \pi \mathrm{i}} \cdot
  \Phi_m^{(m-1-a)}(\tau) ,
\end{gather}
and
from the standard method using the Poisson summation formula
we have
\begin{equation}
  \boldsymbol{\Phi}_m(\tau)
  =
  \left(
    \frac{ \  \mathrm{i}  \ }{\tau}
  \right)^{3/2} \,
  \mathbf{M}_m 
  \cdot
  \boldsymbol{\Phi}_m
  \left( -\frac{  \   1  \  }{\tau} \right) ,
\end{equation}
where
\begin{equation*}
  \boldsymbol{\Phi}_m(\tau)
  =
  \begin{pmatrix}
    \Phi_m^{(m-2)}(\tau) \\
    \vdots
    \\
    \Phi_m^{(1)}(\tau) \\
    \Phi_m^{(0)}(\tau) \\
  \end{pmatrix} ,
\end{equation*}
and $\mathbf{M}_m$ is an $(m-1)\times(m-1)$ matrix,
\begin{equation}
  \bigl( \mathbf{M}_m  \bigr)_{1 \leq  a, b \leq m-1}
  =
  \sqrt{\frac{\  2  \  }{m}} \,
  \sin \left( \frac{\  a \, b   \ }{m} \, \pi \right) .
\end{equation}
Remarkable is  that the theta series  defined by
\begin{equation}
  \ch_\lambda^m(\tau) = \frac{\Phi_{m+2}^{(m-\lambda)}(\tau)}{2\, \bigl( \eta(\tau) \bigr)^3} ,
\end{equation}
where $\eta(\tau)$ is the Dedekind $\eta$-function~\eqref{Dedekind_eta},
is  the affine $\widehat{su}(2)_m$ character
(see Examples in Section~\ref{sec_example})~\cite{Kac90,IGMacdo72}.

As studied in Ref.~\citen{LawrZagi99a},
we have interests in the Eichler integral of the modular form
$\boldsymbol{\Phi}_m(\tau)$.
Generally when  the $q$-series
\begin{equation*}
  F(\tau ) =  \sum_{n=1}^\infty a_n \, q^{n} ,
\end{equation*}
is a modular form with weight $k \in \mathbb{Z}_{\geq 2}$,
the Eichler integral defined as $k-1$ integrations of $F(\tau)$ with
respect to $\tau$, or explicitly defined by
\begin{equation*}
  \widetilde{F}(\tau) = \sum_{n=1}^\infty \frac{a_n}{n^{k-1}} \, q^n ,
\end{equation*}
satisfies
\begin{equation}
  \label{Eichler_identity}
  (c \, \tau +d)^{k-2} \cdot \widetilde{F}(\gamma(\tau))
  - \widetilde{F}(\tau)
  =G_\gamma(\tau) ,
\end{equation}
where $\gamma=
\begin{pmatrix}
  a & b \\
  c & d
\end{pmatrix}
\in SL(2;\mathbb{Z})$, and
$G_\gamma(\tau)$ is the period polynomial
\begin{equation*}
  G_\gamma(z)
  =
  \frac{( 2 \, \pi \, \mathrm{i})^{k-1}}{(k-2)!}
  \int_{\gamma^{-1}(\mathrm{i} \, \infty)}^{\mathrm{i} \infty}
  F(\tau) \,
  (z - \tau)^{k-2} \,
  \mathrm{d} \tau .
\end{equation*}
In our case, the modular form $\Phi_m^{(a)}(\tau)$ in
eq.~\eqref{define_modular}
has a half-integral
weight, and above story does not work any more.
But the Eichler integral as an infinite $q$-series can be defined
in a 
naive sense, and we may find that
\begin{equation*}
  \widetilde{\Phi}_m^{(a)}(\tau)
  =
  m \, \sum_{n=0}^\infty
  \chi_{2 m}^{(a)}(n) \,
  q^{\frac{1}{4m}  n^2} ,
\end{equation*}
which is nothing but a definition~\eqref{Eichler_Phi}.
We note  that a prefactor in above definition is  for our convention.
We can regard $\widetilde{\Phi}_m^{(a)}(\tau)$ as the Eichler integral
of the modular form $\Phi_m^{(a)}(\tau)$ with weight $3/2$.

To study a nearly modular property
(see eq.~\eqref{Eichler_identity})
of this Eichler integral of the half-integral weight modular form, we
first
recall a following result.
\begin{prop}
  \label{prop:L_function}
  Let $C_f(n)$ be a periodic function with mean value $0$ and modulus $f$.
  Then we have an  asymptotic expansion  as $t \searrow 0$
  \begin{equation*}
    \sum_{n=1}^\infty C_f(n) \, \mathrm{e}^{-n^2  t}
    \simeq
    \sum_{k=0}^\infty
    L(-2 \, k, C_f) \,
    \frac{(-t)^k}{k!} ,
  \end{equation*}
  where
  $L(k,C_f)$ is the $L$-function associated with $C_f(n)$,
  and is given by
  \begin{equation*}
    L(-k, C_f)
    =
    - \frac{f^k}{k+1} \,
    \sum_{n=1}^f
    C_f(n) \, B_{k+1} \left( \frac{\ n \ }{f} \right) .
  \end{equation*}
\end{prop}

\begin{proof}
  It is a standard result using the Mellin transformation.
  See, \emph{e.g.}, Ref.~\citen{LawrZagi99a}.
\end{proof}

Using this property,
we obtain the Eichler integral $\widetilde{\Phi}_m^{(a)}(\tau)$
near at a  root of unity as follows.
\begin{prop}
  \label{prop:Eichler_omega}
  The Eichler integral~\eqref{Eichler_Phi} for
  $\tau=\frac{M}{N} \in \mathbb{Q}$ 
  ($N >0$, and $M$, $N$ are coprime integers)
  reduces to
  \begin{equation}
    \label{Eichler_omega}
    \widetilde{\Phi}_m^{(a)}
    \left(\frac{M}{N}\right)
    =
    m \,
    \sum_{n=0}^{m N}
    \chi_{2 m}^{(a)}(n) \,
    \left( 1 - \frac{n}{m \, N} \right) \,
    \mathrm{e}^{\frac{n^2}{2 m N} M \pi \mathrm{i}} .
  \end{equation}
\end{prop}

\begin{proof}
  We have from eq.~\eqref{Eichler_Phi}
  \begin{equation*}
    \widetilde{\Phi}_m^{(a)}
    \left(
      \frac{M}{N} + \mathrm{i} \, \frac{t}{2 \,  \pi}
    \right)
    =
    m \sum_{n=0}^\infty
    C_{2 m N}(n) \, \mathrm{e}^{- \frac{n^2}{4 m} t} ,
  \end{equation*}
  where
  \begin{equation*}
    C_{2 m N}(n)
    =\chi_{2 m}^{(a)} (n) \, \mathrm{e}^{ \frac{M n^2}{2 m N} \pi
    \mathrm{i}} .
  \end{equation*}
  We see that
  $C_{2 m N}(n+2 \, m \, N) = C_{2 m N}(n)$ and
  $C_{2 m N}(2 \, m \, N -n ) = - C_{2 m N}(n)$, and we can apply
  Prop.~\ref{prop:L_function} to get an asymptotic expansion in
  $t   \searrow 0$ as
  \begin{equation*}
    \widetilde{\Phi}_m^{(a)}
    \left( \frac{M}{N} + \mathrm{i} \, \frac{t}{2 \, \pi} \right)
    \simeq
    \sum_{k=0}^\infty
    \frac{L(-2 \, k , C_{2 m N})}{k!} \,
    \left(
      - \frac{t}{4 \, m}
    \right)^k .
  \end{equation*}
  Then we obtain a limiting value
  \begin{align*}
    \widetilde{\Phi}_m^{(a)}
    \left(\frac{M}{N}\right)
    & = 
    m \cdot L(0, C_{2 m N}) .
  \end{align*}
  Recalling an explicit form of the Bernoulli polynomial
  $B_1(x) = x - \frac{1}{2}$ and
  a property
  $C_{2 m N}(2 m N -n)= - C_{2 m N}(n)$,
  we obtain eq.~\eqref{Eichler_omega}.  
\end{proof}

\begin{conj}
  \label{conj:2}
  Let $\widetilde{\Phi}_m^{(a)}(\tau)$ be the Eichler
  integral defined by eq.~\eqref{Eichler_Phi} or~\eqref{define_Phi_m}.
  When $q$ is the $N$-th root of unity,
  $\widetilde{\Phi}_m^{(a)}(1/N)$, which was computed as
  eq.~\eqref{Eichler_omega}, coincides with an
  expression~\eqref{define_Y_a} up to constant,
  \emph{i.e.},
  \begin{equation}
    \label{conjecture_2}
    \widetilde{\Phi}_m^{(a)}
    \left(\frac{1}{N}\right)
    =
    \mathrm{e}^{\frac{(m-1-a)^2}{2 m N} \pi \mathrm{i}} \cdot
    Y_m^{(a)}(\omega) .
  \end{equation}
\end{conj}


\begin{proof}[Proof for a case of $a=0$]
As a case of $a=0$ is related to Kashaev's invariant for the torus
link as in eq.~\eqref{Invariant_and_Y},
this case can be directly  proved by using
eq.~\eqref{Other_Expression_Invariant} as follows;
\begin{align*}
  \mathrm{e}^{\frac{(m-1)^2}{2 m N} \pi \mathrm{i}} \,
  Y_m^{(0)}(\omega)
  & =
  \frac{1}{N} \,\mathrm{e}^{\frac{(m-1)^2}{2 m N} \pi \mathrm{i}} \,
  \langle T(2,2 \, m) \rangle_N
  \\
  & =
  -
  \frac{1}{4 \, m \, N^2}
  \sum_{k=0}^{2 m N}
  k^2 \, \chi_{2 m}^{(0)}(k) \,
  \mathrm{e}^{\frac{k^2}{2 m N} \pi \mathrm{i}}
  \\
  & =
  \sum_{k=0}^{2 m N} \chi_{2 m}^{(0)}(k) \,
  \left(
    \frac{m}{2} - \frac{k}{2 \, N}
  \right) \,
  \mathrm{e}^{\frac{k^2}{2 m N} \pi \mathrm{i}}
  \\
  & =
  \sum_{k=0}^{m N} \chi_{2 m}^{(0)}(k) \,
  \left(
    m - \frac{k}{N}
  \right) \,
  \mathrm{e}^{\frac{k^2}{2 m N} \pi \mathrm{i}}
  =
  \widetilde{\Phi}_m^{(0)} \left( \frac{1}{N} \right) .
\end{align*}
In the third equality, we have summed an expression with
$k\to 2 \, m \, N -k$.
As a result of eq.~\eqref{Eichler_omega}
the statement of Conjecture is true for $a=0$.
\end{proof}

We now  discuss  how
Conjectures~\ref{conj:1} follows from
Conjecture~\ref{conj:2}.
We first recall from eq.~\eqref{Eichler_omega} that
for $N \in \mathbb{Z}$ 
\begin{equation}
  \label{Eichler_integer}
  \widetilde{\Phi}_m^{(a)}(N)
  =
  (1+a) \,
  \mathrm{e}^{\frac{(m-1-a)^2}{2 m} \pi \mathrm{i} N} .
\end{equation}
Following Ref.~\citen{LawrZagi99a} 
(see also Ref.~\citen{Zweg02Thesis}) we define
the period function
\begin{equation}
  r_m^{(a)}(z; \alpha)
   =
  \sqrt{
    \frac{m}{ \  8 \, \mathrm{i}  \  }
  } \,
  \int_\alpha^\infty
  \frac{\Phi_m^{(a)}(\tau)}{\sqrt{\tau -z}} \, \mathrm{d} \tau ,
\end{equation}
where $\alpha \in \mathbb{Q}$.
It is defined for $z$  in the lower half plane,
$z \in \mathbb{H}^-$, but it is analytically continued to
$\mathbb{R}=\partial \mathbb{H}^-$.
To see a modular property of
$\widetilde{\Phi}_m^{(a)}(\alpha + \mathrm{i} \, y)$
in $y \searrow 0$,
we further define
\begin{equation}
  \widehat{\Phi}_m^{(a)}(z)
  =
  \sqrt{
    \frac{m}{\  8 \, \mathrm{i} \ }
  } \,
  \int_{z^*}^\infty
  \frac{\Phi_m^{(a)}(\tau)}{\sqrt{\tau -z}} \, \mathrm{d} \tau ,
\end{equation}
where $z \in \mathbb{H}^-$.
We can find
that this function is  nearly modular of weight $1/2$ by
\begin{align}
  \sum_{b=1}^{m-1}
  \bigl(\mathbf{M}_m \bigr)_{a,b} \,
  \widehat{\Phi}_m^{(m-1-b)}
  \left(-\frac{ \  1 \  }{z} \right)
  & =
  \sum_{b=1}^{m-1}
  \bigl(\mathbf{M}_m \bigr)_{a,b} \,
  \sqrt{
    \frac{m}{\  8 \, \mathrm{i} \ }
  } \,
  \int_{z^*}^0
  \frac{\Phi_m^{(m-1-b)}(-1/s)}{\sqrt{-s^{-1}+z^{-1}}} \,
  \frac{\mathrm{d} s}{s^2}
  \nonumber
  \\
  & =
  \sqrt{\  \mathrm{i} \, z \  } \,
  \left(
    \widehat{\Phi}_m^{(m-1-a)}(z)
    -
    r_m^{(m-1-a)}(z;0)
  \right) .
  \label{modular_hat}
\end{align}

On the other hand,
we have for $z = \alpha + \mathrm{i} \, y $
\begin{align}
  \widehat{\Phi}_m^{(a)}(z)
  & =
  \sqrt{\frac{m}{ \  8 \, \mathrm{i} \  }}
  \sum_{n \in \mathbb{Z}}
  n \, \chi_{2 m}^{(a)}(n)
  \int_{z^*}^\infty
  \frac{
    \mathrm{e}^{\frac{n^2}{2 m} \pi \mathrm{i} \tau}
  }{\sqrt{\tau - z}}
  \, \mathrm{d} \tau
  \nonumber
  \\
  & =
  m \, \sum_{n=1}^\infty
  \chi_{4 m}^{(a)}(n) \, \mathrm{e}^{\frac{n^2}{2 m} \pi \mathrm{i} z}
  \,
  \erfc\left(
    n \, \sqrt{-\frac{\pi \, y}{m}}
  \right) ,
  \nonumber
\end{align}
where
$\erfc(x)$ is the complementary error function
\begin{equation*}
  \erfc(x)
  =
  \frac{2}{\sqrt{\pi}} \int_x^\infty \mathrm{e}^{-t^2} \mathrm{d} t .
\end{equation*}
As $\erfc(0)=1$ and
we know eq.~\eqref{Eichler_omega}
from a definition of $\widetilde{\Phi}_m^{(a)}(\alpha)$,
we get
\begin{equation}
  \widehat{\Phi}_m^{(a)}(\alpha) = \widetilde{\Phi}_m^{(a)}(\alpha) ,
\end{equation}
for $\alpha \in \mathbb{Q}$.
We stress that LHS is a limit from 
the lower half plane $\mathbb{H}^-$ while RHS is
analytically continued
from the upper half plane $\mathbb{H}$.

Taking a limit  $z\to 1/N$ for $N\in \mathbb{Z}$ in
eq.~\eqref{modular_hat}, we obtain
\begin{equation}
  \widehat{\Phi}_m^{(m-1-a)}\left(\frac{1}{N}\right)
  =
  \sqrt{- \mathrm{i} \, N} \,
  \sum_{b=1}^{m-1}
  \bigl(\mathbf{M}_m \bigr)_{a,b} \,
  \widehat{\Phi}_m^{(m-1-b)}
  \left(-N \right)
  +
  r_m^{(m-1-a)}\left(\frac{1}{N};0\right) .
\end{equation}
When we
recall eq.~\eqref{Eichler_integer} and
use  an asymptotic expansion of
$r_m^{(a)}\left(\frac{1}{N};0\right)$ in $N\to\infty$,
we obtain a following proposition.

\begin{prop}
  Let  the Eichler integral
  $\widetilde{\Phi}_m^{(a)}(\tau)$ be defined by
  eq.~\eqref{Eichler_Phi}.
  Asymptotic expansion in $N \to \infty$
  is then given by
  \begin{multline}
    \widetilde{\Phi}_m^{(a)}\left(\frac{1}{N}\right)
    \simeq
    \sqrt{- \mathrm{i} \, N} \,
    \sum_{b=1}^{m-1}
    \sqrt{\frac{2}{\   m \  }} \,
    (m-b)  \,
    \sin
    \left( \frac{(m-1-a) \,b}{m} \, \pi \right) \,
    \mathrm{e}^{-\frac{b^2}{2 m} \pi \mathrm{i} N}
    \\
    +
    \sum_{k=0}^\infty
    \frac{E_k^{(m;a)}}{k!} \,
    \left(
      \frac{\pi \mathrm{i}}{2 \, m \, N}
    \right)^k .
  \end{multline}
\end{prop}

Based on this proposition,
we get a conjecture~\eqref{conjecture_1}
assuming a conjecture~\eqref{conjecture_2}.


\section{Examples}
\label{sec_example}

\subsection{\mathversion{bold}
  ($2,4$)-Torus Link: $m=2$}

\begin{itemize}

\item $q$-series:

  The modular form with weight $3/2$ is
  \begin{align}
    \Phi_2^{(0)}(\tau)
    & =
    \sum_{n \in \mathbb{Z}}
    n \, \chi_4^{(0)}(n) \, q^{\frac{1}{8} n^2}
     =
    \sum_{n \in \mathbb{Z}}
    (-1)^n \, (2 \, n+1) \, q^{\frac{1}{2} (n^2 + n+\frac{1}{4})}
    \\
    & =
    2 \, q^{\frac{1}{8}} \,
    \left(
      1 - 3 \, q + 5 \, q^3 - 7 \, q^6 + 9  \, q^{10} 
      - 11 \, q^{15} + \cdots
    \right) ,
    \nonumber
  \end{align}
  where
  $\chi_4^{(0)}(n)$ is the primitive  character modulo $4$,
  \begin{equation*}
    \begin{array}{c|ccc}
      n \mod 4 & 1 & 3 & \text{others}
      \\
      \hline
      \chi_4^{(0)}(n) & 1 & -1 & 0
    \end{array}
  \end{equation*}
  It is known by Jacobi that we can write
  \begin{equation}
    \Phi_2^{(0)}(\tau)
    =
    2 \,
    \bigl(
      \eta(\tau)
    \bigr)^3 ,
  \end{equation}
  where
  $\eta(\tau)$ is the Dedekind $\eta$-function,
  \begin{equation}
    \label{Dedekind_eta}
    \eta(\tau)
    =
    q^{1/24} \cdot
    (q)_\infty .
  \end{equation}
  From a modular property of the $\eta$-function we see that
  \begin{align}
    \Phi_2^{(0)}(\tau+1)
    & =
    \mathrm{e}^{\frac{1}{4} \pi \mathrm{i}} \,
    \Phi_2^{(0)}(\tau) ,
    \\[2mm]
    \Phi_2^{(0)}(-1/\tau)
    & =
    \left(
      \frac{\ \tau \ }{\mathrm{i}}
    \right)^{3/2} \,
    \Phi_2^{(0)}(\tau) .
  \end{align}

  The Eichler integral is given by
  \begin{align}
    \widetilde{\Phi}_2^{(0)}(\tau)
    & =
    2 \,
    \sum_{n=0}^\infty
    \chi_4^{(0)}(n) \, q^{\frac{1}{8} n^2}
    =
    2 \, q^{\frac{1}{8}} \,
    \sum_{k=0}^\infty
    (-1)^k \, q^{\frac{1}{2} k (k+1)}
    \label{Eichler_4_0}
    \\
    & = 
    2 \, q^{\frac{1}{8}} \,
    \left(
      1 -q + q^3 - q^6 + q^{10} - q^{15} + \cdots
    \right) .
    \nonumber
  \end{align}

\item  root of unity:

  As a limit of $q$ being the $N$-th root of unity, the Eichler
  integral~\eqref{Eichler_4_0} coincides with Kashaev's invariant for
  torus link,
  \begin{equation}
    \widetilde{\Phi}_2^{(0)}
    \left(\frac{1}{N} \right)
    =
    \mathrm{e}^{\frac{\pi \mathrm{i}}{4 N}} \,
    \sum_{c=0}^{N-1}
    (-1)^c \, \omega^{\frac{1}{2} c (c+1)} 
    =
    \frac{1}{N} \,
    \mathrm{e}^{\frac{\pi \mathrm{i}}{4 N}} \,
    \langle T(2,4) \rangle_N .
  \end{equation}
  We see that the Eichler integral~\eqref{Eichler_4_0}
  with $q$ being the $N$-th
  root of unity takes a same form with the original Eichler integral
  up to constant,
  only an infinite sum reduces to a finite sum.
  Using
  $\widetilde{\Phi}_2^{(0)}(N)=\mathrm{e}^{\pi \mathrm{i} N/4}$, we have
  the nearly modular property
  (see eq.~\eqref{torus_link_asymptotic}),
  \begin{equation}
    \widetilde{\Phi}_2^{(0)}\left(\frac{1}{N}\right)
    \simeq
    \sqrt{- \mathrm{i} \, N} \,
    \widetilde{\Phi}_2^{(0)}(-N)
    +
    \sum_{n=0}^\infty
    \frac{E_n^{(2;0)}}{n!} \,
    \left(
      \frac{\pi \, \mathrm{i}}{4 \, N}
    \right)^n ,
  \end{equation}
  where $E_n^{(2;0)}$ is the Euler number defined by
  \begin{gather*}
    E_n^{(2;0)}
    =
    -\frac{2^{4 n+1}}{2 \, n+1}
    \left(
      B_{2 n +1} \left(\frac{1}{4}\right)
      - B_{2 n +1}\left(\frac{3}{4}\right)
    \right) ,
  \end{gather*}
  some of which are  given as
  \begin{align*}
    \frac{1}{\ch(x) }
    & =
    \sum_{n=0}^\infty
    \frac{E_n^{(2;0)}}{  \   (2 \, n) !   \  } \ x^{2n}
    \\
    &   =
    1 - \frac{1}{2} \, x^2 +
    \frac{5}{24} \, x^4 - \frac{61}{720} \, x^6 + \cdots .
  \end{align*}

\end{itemize}

\subsection{\mathversion{bold}
  ($2,6$)-Torus Link: $m=3$}

\begin{itemize}
\item 
  $q$-series:

  A set of the theta series is given by
  \begin{subequations}
    \begin{align}
      \Phi_3^{(0)}(\tau) & =
      \sum_{n \in \mathbb{Z}} n \, \chi_6^{(0)}(n) \,
      q^{\frac{1}{12} n^2}
      \\
      & =
      4 \, q^{\frac{1}{3}} \,
      \left(
        1 - 2 \, q + 4 \, q^5 - 5 \, q^8 + 7 \, q^{16} - 8 \, q^{21}
        + \cdots
      \right) ,
      \nonumber
      \\[2mm]
      \Phi_3^{(1)}(\tau) & =
      \sum_{n \in \mathbb{Z}} n \, \chi_6^{(1)}(n) \,
      q^{\frac{1}{12} n^2}
      \\
      &
      =
      2 \, q^{\frac{1}{12}} \,
      \left(
        1 -  5 \, q^2 + 7 \, q^4 - 11 \, q^{10} + 13 \, q^{14}
        - 17 \, q^{24} + \cdots
      \right) ,
      \nonumber
    \end{align}
  \end{subequations}
  where
  \begin{align*}
    &
    \begin{array}{c|ccc}
      n \mod 6 & 2 & 4 & \text{others}
      \\
      \hline
      \chi_6^{(0)}(n) & 1 & -1 & 0
    \end{array}
    & &
    \begin{array}{c|ccc}
      n \mod 6 & 1 & 5 & \text{others}
      \\
      \hline
      \chi_6^{(1)}(n) & 1 & -1 & 0
    \end{array}
  \end{align*}
  We note~\cite{IGMacdo72} that
  these series can  be  written in terms of the Dedekind
  $\eta$-function as
  \begin{subequations}
    \begin{align}
      \Phi_3^{(0)}(\tau)
      & =
      4 \, \mathrm{e}^{\frac{1}{12} \pi \mathrm{i}} \,
      \frac{
        \bigl( \eta(2 \, \tau  ) \bigr)^5
      }{
        \bigl( \eta ( \tau + \frac{1}{2} ) \bigr)^2
      }
      =
      4 \,
      \frac{
        \bigl( \eta(\tau) \, \eta(4 \, \tau) \bigr)^2
      }{
        \eta(2 \, \tau)
      } ,
      \\[2mm]
      \Phi_3^{(1)}(\tau)
      & =
      2 \,
      \frac{
        \bigl( \eta(2 \, \tau) \bigr)^5
      }{
        \bigl( \eta (4 \, \tau) \bigr)^2
      } .
    \end{align}
  \end{subequations}
  The modular property is written as
  \begin{gather}
    \begin{pmatrix}
      \Phi_3^{(1)}(\tau+1) \\[2mm]
      \Phi_3^{(0)}(\tau+1) 
    \end{pmatrix}
    =
    \begin{pmatrix}
      \mathrm{e}^{\frac{1}{6} \pi \mathrm{i}} & 0
      \\[2mm]
      0 & \mathrm{e}^{\frac{2}{3} \pi \mathrm{i}} 
    \end{pmatrix}
    \,
    \begin{pmatrix}
      \Phi_3^{(1)}(\tau) \\[2mm]
      \Phi_3^{(0)}(\tau)
    \end{pmatrix} ,
    \\[2mm]
    \begin{pmatrix}
      \Phi_3^{(1)}(-1/\tau) \\[2mm]
      \Phi_3^{(0)}(-1/\tau) 
    \end{pmatrix}
    =
    \left(\frac{\   \tau   \   }{\mathrm{i}}\right)^{3/2}
    \cdot
    \frac{1}{\sqrt{2}}
    \begin{pmatrix}
      1 & 1
      \\[2mm]
      1 & -1
    \end{pmatrix}
    \cdot
    \begin{pmatrix}
      \Phi_3^{(1)}(\tau) \\[2mm]
      \Phi_3^{(0)}(\tau)
    \end{pmatrix} .
  \end{gather}

  The Eichler integrals are then defined by
  \begin{subequations}
    \label{Eichler_3}
    \begin{align}
      \widetilde{\Phi}_3^{(0)}(\tau)
      & =
      3 \,
      q^{ \frac{1}{3}} \,
      \sum_{a=0}^\infty
      (-1)^a \, q^{\frac{1}{2} a(a+1)} 
      \sum_{b=0}^a
      q^{b (b+1)}
      \begin{bmatrix}
        a  \\ b
      \end{bmatrix}
      \nonumber 
      \\
      & =
      3 \, 
      \sum_{n=0}^\infty
      \chi_6^{(0)}(n) \, q^{\frac{1}{12} n^2}
      \label{Eichler_3_0}
      \\
      & =
      3 \,    q^{\frac{1}{3}} \,
      \left(
        1 -q + q^5 - q^8 + q^{16} - q^{21} + \cdots
      \right) ,
      \nonumber
      \\[2mm]
      \widetilde{\Phi}_3^{(1)}(\tau)
      &=
      3 \,
      q^{ \frac{1}{12}} \,
      \sum_{a=0}^\infty
      (-1)^a \, q^{\frac{1}{2} a(a+1)} 
      \sum_{b=0}^{a+1}
      q^{b^2}
      \begin{bmatrix}
        a+1  \\ b
      \end{bmatrix}
      \nonumber
      \\
      & =
      3 \, \sum_{n=0}^\infty
      \chi_6^{(1)}(n) \, q^{\frac{1}{12} n^2}
      \label{Eichler_3_1}
      \\
      & =
      3 \,    q^{\frac{1}{12}} \,
      \left(
        1 -q^2 + q^4 - q^{10} + q^{14} - q^{24} + \cdots
      \right) .
      \nonumber
    \end{align}
  \end{subequations}
  We note that Zagier's identity~\cite{DZagie01a} leads us to find
  \begin{equation*}
    \widetilde{\Phi}_3^{(1)}(\tau)
    =
    3 \,  q^{\frac{1}{12}}
    \sum_{n=0}^\infty
    (-1)^n \, (-1 ; q^2)_{n+1}  .
  \end{equation*}

\item  root of unity:

  From Prop.~\ref{prop:Eichler_omega},
  the Eichler integrals~\eqref{Eichler_3}
  reduce  in a case of $q$ being $N$-th root of unity to
  \begin{equation}
    \widetilde{\Phi}_3^{(a)}
    \left( \frac{1}{N} \right)
    =
    3 \,
    \sum_{n=0}^{3 N}
    \chi_{6}^{(a)}(n) \,
    \left(
      1 - \frac{n}{3 \, N}
    \right) \,
    \mathrm{e}^{\frac{n^2}{6 N} \pi \mathrm{i}} .
  \end{equation}
  This coincides with
  \begin{subequations}
    \label{conjecture_for_3}
    \begin{align}
      \widetilde{\Phi}_3^{(0)}\left(\frac{1}{N}\right)
      & =
      \mathrm{e}^{\frac{2 \pi \mathrm{i}}{3 N}} \,
      \sum_{a,b =0}^{N-1}
      (-1)^a \, 
      \omega^{\frac{1}{2} a (a+1) + b (b+1)} \,
      \begin{bmatrix}
        a \\ b
      \end{bmatrix}
      =
      \frac{1}{N} \,
      \mathrm{e}^{\frac{2 \pi \mathrm{i}}{3 N}} \,
      \langle T(2,6) \rangle_N ,
      \\[2mm]
      \widetilde{\Phi}_3^{(1)}\left(\frac{1}{N}\right)
      & =
      \mathrm{e}^{\frac{ \pi \mathrm{i}}{6 N}} \,
      \sum_{a,b =0}^{N-1}
      (-1)^a \, 
      \omega^{\frac{1}{2} a (a+1) + b^2} \,
      \begin{bmatrix}
        a+1 \\ b
      \end{bmatrix} .
    \end{align}
  \end{subequations}
  where
  the second identity remains to be proved
  (the first identity was established from an asymptotic expansion
  due to a discussion in the
  previous section).
  We have checked numerically  a validity of this identity.

  As we have
  \begin{equation*}
    \begin{pmatrix}
      \widetilde{\Phi}_3^{(1)}(N) \\[2mm]
      \widetilde{\Phi}_3^{(0)}(N)
    \end{pmatrix}
    =
    \begin{pmatrix}
      2 \, \mathrm{e}^{\frac{1}{6} \pi \mathrm{i} N}
      \\[2mm]
      \mathrm{e}^{\frac{2}{3} \pi \mathrm{i} N}
    \end{pmatrix} ,
  \end{equation*}
  modular property and eqs.~\eqref{conjecture_for_3} supports that
  \begin{equation}
    \begin{pmatrix}
      \widetilde{\Phi}_3^{(1)}\left(\frac{1}{N}\right) \\[2mm]
      \widetilde{\Phi}_3^{(0)}\left(\frac{1}{N}\right)
    \end{pmatrix}
    \simeq
    \sqrt{ - \mathrm{i} \, N} \cdot
    \frac{1}{\sqrt{2}}
    \begin{pmatrix}
      1 & 1 \\[2mm]
      1 & -1
    \end{pmatrix}
    \cdot
    \begin{pmatrix}
      \widetilde{\Phi}_3^{(1)}(-N) \\[2mm]
      \widetilde{\Phi}_3^{(0)}(-N)
    \end{pmatrix}
    +
    \sum_{n=0}^\infty
    \frac{1}{\   n !  \ }
    \begin{pmatrix}
      E_n^{(3:1)} \\[2mm]
      E_n^{(3:0)}
    \end{pmatrix}
    \,
    \left(
      \frac{ \pi \, \mathrm{i}}{6 \, N}
    \right)^n ,
  \end{equation}
  where the generalized Euler number is defined by
  \begin{equation*}
    E_n^{(3;a)}
    =
    -\frac{3 \cdot 6^{2 n}}{2 \, n+1} \,
    \left(
      B_{2 n+1}
      \left( \frac{2-a}{6} \right)
      -
      B_{2 n+1}
      \left( \frac{4+a}{6} \right)
    \right) ,
  \end{equation*}
  or some of them are as follows;
  \begin{align*}
    \frac{3}{\sh( 3 \, x)}
    \begin{pmatrix}
      \sh(2 \, x) 
      \\[2mm]
      \sh ( x)
    \end{pmatrix}
    & =
    \sum_{k=0}^\infty
    \begin{pmatrix}
      E_k^{(3;1)}
      \\[2mm]
      E_k^{(3;0)}
    \end{pmatrix}
    \,
    \frac{x^{2 k}}{(2 \, k)!} 
    \\
    & =
    \begin{pmatrix}
      2 \\[2mm]
      1
    \end{pmatrix}
    -
    \begin{pmatrix}
      \frac{10}{3} \\[2mm]
      \frac{8}{3} 
    \end{pmatrix}
    \, \frac{x^2}{2}
    +
    \begin{pmatrix}
      34 \\[2mm]
      32
    \end{pmatrix}
    \, \frac{x^4}{24}
    -
    \begin{pmatrix}
      910 \\[2mm]
      896
    \end{pmatrix}
    \, \frac{x^6}{720}
    + \cdots .
  \end{align*}
\end{itemize}

\subsection{\mathversion{bold}
  ($2,8$)-Torus Link: $m=4$}

\begin{itemize}
\item $q$-series:
  
  A set of theta series is given by
  \begin{subequations}
    \begin{align}
      \Phi_4^{(0)}(\tau) & =
      \sum_{n \in \mathbb{Z}} n \, \chi_8^{(0)}(n) \,
      q^{\frac{1}{16} n^2}
      \\
      & =
      \left(
        \frac{
          \bigl( \eta(\tau) \bigr)^3
        }{
          \eta \left( \frac{\tau}{2} \right) \, \eta(2 \, \tau)
        }
      \right)^3
      -
      \left(
        \eta \left( \frac{\tau}{2} \right)
      \right)^3
      \nonumber \\
      & =
      2 \, q^{\frac{9}{16}} \,
      \left(
        3 - 5 \, q + 11 \, q^7 - 13 \, q^{10} + 19 \, q^{22} - 21 \, q^{27}
        + \cdots
      \right) ,
      \nonumber
      \\[2mm]
      \Phi_4^{(1)}(\tau) & =
      \sum_{n \in \mathbb{Z}} n \, \chi_8^{(1)}(n) \,
      q^{\frac{1}{16} n^2}
      \\
      & =
      4 \, \bigl( \eta(2 \, \tau) \bigr)^3
      \nonumber \\
      &
      =
      4 \, q^{\frac{1}{4}} \,
      \left(
        1 -  3 \, q^2 + 5 \, q^6 - 7 \, q^{12} + 9 \, q^{20}
        - 11 \, q^{30} + \cdots
      \right) ,
      \nonumber
      \\[2mm]
      \Phi_4^{(2)}(\tau) & =
      \sum_{n \in \mathbb{Z}} n \, \chi_8^{(2)}(n) \,
      q^{\frac{1}{16} n^2}
      \\
      & =
      \left(
        \frac{
          \bigl( \eta(\tau) \bigr)^3
        }{
          \eta ( \frac{\tau}{2}) \, \eta(2 \, \tau)
        }
      \right)^3
      +
      \left(
        \eta \left( \frac{\tau}{2} \right)
      \right)^3
      \nonumber \\
      &
      =
      2 \, q^{\frac{1}{16}} \,
      \left(
        1 -  7 \, q^3 + 9 \, q^5 - 15 \, q^{14} + 17 \, q^{18}
        - 23 \, q^{33} + \cdots
      \right) ,
      \nonumber
    \end{align}
  \end{subequations}
  which are modular coinvariant;
  \begin{gather}
    \begin{pmatrix}
      \Phi_4^{(2)}(\tau+1) \\[2mm]
      \Phi_4^{(1)}(\tau+1) \\[2mm]
      \Phi_4^{(0)}(\tau+1) 
    \end{pmatrix}
    =
    \begin{pmatrix}
      \mathrm{e}^{\frac{1}{8} \pi \mathrm{i}} & 0 & 0
      \\[2mm]
      0 & \mathrm{e}^{\frac{1}{2} \pi \mathrm{i}} & 0
      \\[2mm]
      0 & 0 & \mathrm{e}^{\frac{9}{8} \pi \mathrm{i}} 
    \end{pmatrix}
    \,
    \begin{pmatrix}
      \Phi_4^{(2)}(\tau) \\[2mm]
      \Phi_4^{(1)}(\tau) \\[2mm]
      \Phi_4^{(0)}(\tau)
    \end{pmatrix} ,
    \\[2mm]
    \begin{pmatrix}
      \Phi_4^{(2)}(-1/\tau) \\[2mm]
      \Phi_4^{(1)}(-1/\tau) \\[2mm]
      \Phi_4^{(0)}(-1/\tau) 
    \end{pmatrix}
    =
    \left(\frac{\   \tau   \   }{\mathrm{i}}\right)^{3/2}
    \cdot
    \frac{1}{2}
    \begin{pmatrix}
      1 & \sqrt{2} & 1
      \\[2mm]
      \sqrt{2} & 0 & - \sqrt{2}
      \\[2mm]
      1 & - \sqrt{2} & 1
    \end{pmatrix}
    \cdot
    \begin{pmatrix}
      \Phi_4^{(2)}(\tau) \\[2mm]
      \Phi_4^{(1)}(\tau) \\[2mm]
      \Phi_4^{(0)}(\tau)
    \end{pmatrix} .
  \end{gather}

  We then have the Eichler integral as
  \begin{subequations}
    \begin{align}
      \widetilde{\Phi}_4^{(0)}(\tau)
      & =
      4 \,
      q^{ \frac{9}{16}} \,
      \sum_{a=0}^\infty
      (-1)^a \, q^{\frac{1}{2} a(a+1)} 
      \sum_{b=0}^a
      q^{b (b+1)}
      \begin{bmatrix}
        a  \\ b
      \end{bmatrix}
      \sum_{c=0}^b
      q^{c (c+1)}
      \begin{bmatrix}
        b  \\ c
      \end{bmatrix}
      \nonumber 
      \\
      & =
      4 \, 
      \sum_{n=0}^\infty
      \chi_8^{(0)}(n) \, q^{\frac{1}{16} n^2}
      \\
      & =
      4 \,    q^{\frac{9}{16}} \,
      \left(
        1 -  q +  q^7 -  q^{10} +  q^{22} -  q^{27}
        + \cdots
      \right) ,
      \nonumber
      \\[2mm]
      \widetilde{\Phi}_4^{(1)}(\tau)
      &=
      4 \,
      q^{ \frac{1}{4}} \,
      \sum_{a=0}^\infty
      (-1)^a \, q^{\frac{1}{2} a(a+1)} 
      \sum_{b=0}^{a}
      q^{b(b+1)}
      \begin{bmatrix}
        a \\ b
      \end{bmatrix}
      \sum_{c=0}^{b+1}
      q^{c^2}
      \begin{bmatrix}
        b+1 \\ c
      \end{bmatrix}
      \nonumber
      \\
      & =
      4 \, \sum_{n=0}^\infty
      \chi_8^{(1)}(n) \, q^{\frac{1}{16} n^2}
      \\
      & =
      4 \,    q^{\frac{1}{4}} \,
      \left(
        1 -   q^2 +  q^6 -  q^{12} +  q^{20} -  q^{30} + \cdots
      \right) ,
      \nonumber
      \\[2mm]
      \widetilde{\Phi}_4^{(2)}(\tau)
      &=
      4 \,
      q^{ \frac{1}{16}} \,
      \sum_{a=0}^\infty
      (-1)^a \, q^{\frac{1}{2} a(a+1)} 
      \sum_{b=0}^{a+1}
      q^{b^2}
      \begin{bmatrix}
        a+1 \\ b
      \end{bmatrix}
      \sum_{c=0}^{b}
      q^{c^2}
      \begin{bmatrix}
        b \\ c
      \end{bmatrix}
      \nonumber
      \\
      & =
      4 \, \sum_{n=0}^\infty
      \chi_8^{(2)}(n) \, q^{\frac{1}{16} n^2}
      \\
      & =
      4 \,    q^{\frac{1}{16}} \,
      \left(
        1 -   q^3 +  q^5 -  q^{14} +  q^{18} -  q^{33} + \cdots
      \right) .
      \nonumber
    \end{align}
  \end{subequations}
  One sees that
  \begin{equation*}
    \widetilde{\Phi}_4^{(1)}(\tau)
    =
    2 \,
    \widetilde{\Phi}_2^{(0)}(2 \, \tau) .
  \end{equation*}

\item root of unity:
  
  Limiting value of the Eichler integrals when $q$ goes to the $N$-th
  primitive root of unity  is given by
  \begin{equation}
    \widetilde{\Phi}_4^{(a)}
    \left( \frac{1}{N} \right)
    =
    4 \,
    \sum_{n=0}^{4 N}
    \chi_{8}^{(a)}(n) \,
    \left(
      1 - \frac{n}{4 \, N}
    \right) \,
    \mathrm{e}^{\frac{n^2}{8 N} \pi \mathrm{i}} ,
  \end{equation}
  which is rewritten as
  \begin{subequations}
    \begin{align}
      \label{Eichler_q_4_0}
      \widetilde{\Phi}_4^{(0)}\left(\frac{1}{N}\right)
      & =
      \mathrm{e}^{\frac{9 \pi \mathrm{i}}{8 N}} \,
      \sum_{a,b ,c  =0}^{N-1}
      (-1)^a \, 
      \omega^{\frac{1}{2} a (a+1) + b (b+1)+ c(c+1)} \,
      \begin{bmatrix}
        a \\ b
      \end{bmatrix} \,
      \begin{bmatrix}
        b \\ c
      \end{bmatrix}
      =
      \frac{1}{N} \,
      \mathrm{e}^{\frac{9 \pi \mathrm{i}}{8 N}} \,
      \langle T(2,8) \rangle_N ,
      \\[2mm]
      \widetilde{\Phi}_4^{(1)}\left(\frac{1}{N}\right)
      & =
      \mathrm{e}^{\frac{ \pi \mathrm{i}}{2 N}} \,
      \sum_{a,b,c  =0}^{N-1}
      (-1)^a \, 
      \omega^{\frac{1}{2} a (a+1) + b(b+1) + c^2} \,
      \begin{bmatrix}
        a \\ b
      \end{bmatrix}  \,
      \begin{bmatrix}
        b+1 \\ c
      \end{bmatrix} ,
      \\[2mm]
      \widetilde{\Phi}_4^{(2)}\left(\frac{1}{N}\right)
      & =
      \mathrm{e}^{\frac{ \pi \mathrm{i}}{8 N}} \,
      \sum_{a,b,c  =0}^{N-1}
      (-1)^a \, 
      \omega^{\frac{1}{2} a (a+1) + b^2 + c^2} \,
      \begin{bmatrix}
        a+1 \\ b
      \end{bmatrix}  \,
      \begin{bmatrix}
        b \\ c
      \end{bmatrix} .
    \end{align}
  \end{subequations}
  Though we have  checked these three equalities  numerically,
  we   proved only  eq.~\eqref{Eichler_q_4_0} in this article.

  The nearly modular properties are written as
  \begin{multline}
    \begin{pmatrix}
      \widetilde{\Phi}_4^{(2)}\left(\frac{1}{N}\right) \\[2mm]
      \widetilde{\Phi}_4^{(1)}\left(\frac{1}{N}\right) \\[2mm]
      \widetilde{\Phi}_4^{(0)}\left(\frac{1}{N}\right)
    \end{pmatrix}
    \simeq
    \sqrt{ - \mathrm{i} \, N} \cdot
    \frac{1}{2}
    \begin{pmatrix}
      1 & \sqrt{2} & 1
      \\[2mm]
      \sqrt{2} & 0 & - \sqrt{2}
      \\[2mm]
      1 & - \sqrt{2} & 1
    \end{pmatrix}
    \cdot
    \begin{pmatrix}
      \widetilde{\Phi}_4^{(2)}(-N) \\[2mm]
      \widetilde{\Phi}_4^{(1)}(-N) \\[2mm]
      \widetilde{\Phi}_4^{(0)}(-N)
    \end{pmatrix}
    \\
    +
    \sum_{n=0}^\infty
    \frac{1}{\   n !  \ }
    \begin{pmatrix}
      E_n^{(4:2)} \\[2mm]
      E_n^{(4:1)} \\[2mm]
      E_n^{(4:0)}
    \end{pmatrix}
    \,
    \left(
      \frac{ \pi \, \mathrm{i}}{8 \, N}
    \right)^n ,
  \end{multline}
  where we have
  \begin{equation*}
    \begin{pmatrix}
      \widetilde{\Phi}_4^{(2)}(N) \\[2mm]
      \widetilde{\Phi}_4^{(1)}(N) \\[2mm]
      \widetilde{\Phi}_4^{(0)}(N)
    \end{pmatrix}
    =
    \begin{pmatrix}
      3 \, \mathrm{e}^{\frac{1}{8} \pi \mathrm{i} N}
      \\[2mm]
      2 \, \mathrm{e}^{\frac{1}{2} \pi \mathrm{i} N}
      \\[2mm]
      \mathrm{e}^{\frac{9}{8} \pi \mathrm{i} N}
    \end{pmatrix} ,
  \end{equation*}
  and   the generalized Euler number is defined by
  \begin{equation*}
    E_n^{(4;a)}
    =
    -\frac{ 2^{6 n+2}}{2 \, n+1} \,
    \left(
      B_{2 n+1}
      \left( \frac{3-a}{8} \right)
      -
      B_{2 n+1}
      \left( \frac{5+a}{8} \right)
    \right) .
  \end{equation*}
  Some of them are explicitly given as follows;
  \begin{align*}
    \frac{4}{\sh( 4 \, x)}
    \begin{pmatrix}
      \sh(3 \, x)       \\[2mm]
      \sh(2 \, x)       \\[2mm]
      \sh ( x)
    \end{pmatrix}
    & =
    \sum_{k=0}^\infty
    \begin{pmatrix}
      E_k^{(4;2)}      \\[2mm]
      E_k^{(4;1)}      \\[2mm]
      E_k^{(4;0)}
    \end{pmatrix} \,
    \frac{x^{2 k}}{(2 \, k)!} 
    \\
    & =
    \begin{pmatrix}
      3 \\[2mm]
      2 \\[2mm]
      1
    \end{pmatrix}
    -
    \begin{pmatrix}
      7 \\[2mm]
      8 \\[2mm]
      5
    \end{pmatrix}
    \, \frac{x^2}{2}
    +
    \begin{pmatrix}
      {119} \\[2mm]
      {160} \\[2mm]
      {109} 
    \end{pmatrix}
    \, \frac{x^4}{24}
    -
    \begin{pmatrix}
      {5587} \\[2mm]
      {7808} \\[2mm]
      {5465} 
    \end{pmatrix}
    \, \frac{x^6}{720}
    + \cdots .
  \end{align*}

\end{itemize}

\section*{Acknowledgments}
The author would like to thank Anatol N. Kirillov for suggesting to study
torus links.
He also thanks Hitoshi Murakami for communications
in constructing  quantum knot invariants.
Thanks are to Don Zagier for comments on nearly modular form.
He thanks George Andrews for his interests.
This work is supported in part by the Sumitomo Foundation,
and Grant-in-Aid for Young Scientists
from the Ministry of Education, Culture, Sports, Science and
Technology of Japan.

\bibliographystyle{siam}

\end{document}